\documentclass[twocolumn, journal]{IEEEtran}
\IEEEoverridecommandlockouts

\usepackage{diagbox}
\usepackage{bm}
\usepackage{color}
\usepackage{graphicx}
\usepackage{amsmath}
\usepackage{amssymb}
\usepackage{algorithm}
\usepackage{algorithmic}
\usepackage{ulem}
\usepackage{lineno}
\usepackage{lettrine}
\usepackage{subfigure}
\usepackage{epstopdf}
\usepackage{stfloats}
\usepackage{color}
\usepackage{graphicx}
\usepackage{amsmath}
\usepackage{amssymb}
\usepackage{algorithm}
\usepackage{algorithmic}
\usepackage{amsmath}
\usepackage{multirow}
\usepackage{booktabs}
\usepackage{array}
\usepackage{amsthm}
\usepackage{stfloats}
\usepackage{caption}
\usepackage{bm}
\usepackage{booktabs}
\usepackage{setspace}
\usepackage{makecell}
\usepackage{cases}

\newcommand{\eqdef}{\triangleq}
\newcommand{\be}{\begin{equation}}
\newcommand{\ee}{\end{equation}}
\newcommand{\bea}{\begin{eqnarray}}
\newcommand{\eea}{\end{eqnarray}}
\newcommand{\ba}{\begin{array}}
\newcommand{\ea}{\end{array}}

\title{Joint Beamforming Designs for\\ Active Reconfigurable Intelligent Surface:\\ A Sub-Connected Array Architecture
\thanks{Part of this paper has been presented in the IEEE Sensor Array and Multichannel Signal Processing Workshop (SAM), 2022 \cite{SAM}.}
\thanks{Q. Zhu, M. Li, R. Liu, and Y. Liu are with the School of Information and Communication Engineering, Dalian University of Technology, Dalian 116024, China (e-mail: qzhu@mail.dlut.edu.cn; mli@dlut.edu.cn; liurang@mail.dlut.edu.cn; yangliu{\_}613@dlut.edu.cn).}
\thanks{
Q. Liu is with the School of Computer Science and Technology, Dalian University of Technology, Dalian 116024, China (e-mail: qianliu@dlut.edu.cn).}
}

\author{ \IEEEauthorblockN{Qi Zhu, Ming Li, \textit{Senior Member, IEEE}, Rang Liu, \textit{Graduate Student Member, IEEE},\\ Yang Liu, \textit{Member, IEEE}, and Qian Liu, \textit{Member, IEEE}}}

\begin{document}
\maketitle
\pagestyle{empty}
\begin{abstract}
  Reconfigurable intelligent surface (RIS) is regarded as a promising technology with great potential to boost wireless networks. Affected by the ``double fading" effect, however, conventional passive RIS cannot bring considerable performance improvement when users are not close enough to RIS. Recently, active RIS is introduced to combat the double fading effect by actively amplifying incident signals with the aid of integrated reflection-type amplifiers. In order to reduce the hardware cost and energy consumption due to massive active components in the conventional fully-connected active RIS, a novel hardware-and-energy efficient sub-connected active RIS architecture has been proposed recently, in which multiple reconfigurable electromagnetic elements are driven by only one amplifier. In this paper, we first develop an improved and accurate signal model for the sub-connected active RIS architecture. Then, we investigate the joint transmit precoding and RIS reflection beamforming (i.e., the reflection phase-shift and amplification coefficients) designs in multiuser multiple-input single-output (MU-MISO) communication systems. Both sum-rate maximization and power minimization problems are solved by leveraging fractional programming (FP), block coordinate descent (BCD), second-order cone programming (SOCP), alternating direction method of multipliers (ADMM), and majorization-minimization (MM) methods. Extensive simulation results verify that compared with the conventional fully-connected structure, the proposed sub-connected active RIS can significantly reduce the hardware cost and power consumption, and achieve great performance improvement when power budget at RIS is limited.
\end{abstract}


\section{Introduction}
Over the past decades, wireless communications have been constantly undergoing tremendous changes. As the fifth-generation (5G) mobile network is becoming a commercial reality, researchers are paying more attention to the sixth-generation (6G) communication. Recently emerging reconfigurable intelligent surface (RIS) technology is deemed as a promising key enabler in 6G networks owing to its superior capability of intelligently reconfiguring wireless communication environment \cite{Pan}, \cite{GC}.
Generally speaking, an RIS is a planar array composed of passive electromagnetic elements, each of which can independently tune the phase-shift and amplitude of the incident signal \cite{Basar}-\cite{Wu}. Benefiting from this ability, the reflected signal can be properly adjusted to enhance the signal strength at the receiver. This superior capability means that RIS can adaptively manipulate wireless propagation to fundamentally tackle the blockage issue and introduce additional degrees of freedom (DoFs) to improve the communication performance.

Attracted by its sheer advantages, the applications of RIS in various wireless communication scenarios have been extensively investigated. A majority of research efforts have been devoted to the RIS designs for spectral/energy efficiency maximization \cite{Zhou}, \cite{Huang}, sum-rate maximization \cite{Zhou2}, \cite{WangP}, transmit power minimization \cite{Wu3}, \cite{Wu4}, etc. Moreover, RIS is often integrated with other emerging technologies to unlock additional potentials, such as physical-layer security (PLS) \cite{CuiM}, simultaneous wireless information and power transfer (SWIPT) \cite{Wu5}, wideband orthogonal frequency division multiplexing (OFDM) \cite{Li2}, symbol-level precoding \cite{LiuR}, \cite{LiuR2}, and integrated sensing and communications (ISAC) \cite{WangX}, \cite{LiuR3}.

With the proliferation of researches on RIS, a fatal ``double fading" effect has received a great deal of attention \cite{Long}, \cite{Zhang}. The so-called ``double fading'' effect is essentially a kind of multiplicative fading, caused by the fact that the signal reflected by RIS has to pass through a cascade link consisting of the transmitter-RIS channel and the RIS-receiver channel. In other words, the reflected signal suffers from large-scale fading twice. As a result, the performance improvement achieved by the RIS-assisted reflection channel is marginal when users are not close enough to RIS. Traditionally, this issue can be tackled by simply increasing the number of RIS elements. However, massive RIS elements will lead to unaffordable pilot overhead for channel estimation \cite{HuC} and real-time beamforming design challenges \cite{PanC}. Moreover, the increase in physical size will restrict the placement of RIS \cite{Najafi}. Inevitably, the ``double fading" effect becomes the major hurdle for the practical deployment of conventional passive RIS.

In order to overcome the aforementioned defect of passive RIS, the concept of active RIS has been introduced from the latest literature \cite{Long}, \cite{Zhang}. While the novel active RIS still has the ability to reflect the incoming signals with adjusted phase-shifts, each element of active RIS integrates an additional reflection-type amplifier to actively amplify the signal, which can be realized by many existing active components \cite{Bousquet}-\cite{KuoC}. Therefore, active RIS can effectively conquer the ``double fading" effect by amplifying the weak signal propagating through the transmitter-RIS channel.

Recently, some researches have focused on exploiting the advantages of active RIS \cite{Long}, \cite{Zhang}, \cite{You}-\cite{GeY}. In \cite{Long}, the authors for the first time proposed the concept of active RIS and verified the advantages of the active RIS aided system compared to the passive one in terms of received signal-to-noise ratio (SNR) with simulation results. The authors in \cite{Zhang} demonstrated that the active RIS exhibits a 67\% sum-rate gain compared to the typical no-RIS system, while the conventional passive RIS can realize only a negligible gain of about 3\%. In \cite{You}, the authors studied the optimal placement of active RIS in the downlink and/or uplink communication. The authors in \cite{Khoshafa} and \cite{Dong} employed the active RIS in wireless communication systems to enhance the security performance, where secrecy outage probability and security rate are utilized as metrics, respectively. Active RIS-aided SWIPT system was studied in \cite{GaoY}, where an active RIS was deployed to assist an access point (AP) to convey information and energy simultaneously to information users and energy users for significant performance enhancement. In addition, active RIS-assisted wireless powered communication network (WPCN) and unmanned aerial vehicles (UAV) secure communication were investigated in \cite{ZengP} and \cite{GeY}, respectively.

While active RIS has been attracting more attention, it is worth noting that the fully-connected active RIS structure, in which each element is equipped with a dedicated amplifier, is uneconomical due to the expensive hardware cost and high power consumption of massive active components. To further lower the cost of active RIS, several new structural designs are recently introduced in \cite{Nguyen}-\cite{KLiu}. The concept of hybrid active-passive RIS was presented in \cite{Nguyen}, \cite{Nguyen2}. The key idea of hybrid RIS is to add a few active elements to the traditional passive RIS, enabling it to reflect and amplify incident signals simultaneously. However, the hybrid RIS structure allows only a few elements to amplify the incident signals, which severely restricts the performance improvement. In the very recent research \cite{KLiu}, the authors proposed a novel realization of active RIS, in which multiple reflection elements were grouped together and controlled by the same amplifier circuit to lower the energy and hardware expenditure. Nevertheless, the signal model of this pioneering structure is relatively simple and rough, in which the process of power combination and re-distribution due to multiple reflection elements sharing an amplifier is not accurately described.
It is worth noting that the sub-connected array architecture of active RIS is fundamentally different from the element grouping strategy of passive RIS \cite{MaoZ}-\cite{YanY} for following two main reasons. First, the motivations of element grouping are different. While the former's goal is to reduce the power consumption and hardware cost, the latter aims to decrease the overhead of channel estimation and simplify the passive beamforming design. Second, the structures and mechanisms of element grouping are different. In the sub-connected active RIS, the elements in the same group are connected to and driven by the same amplifier, i.e., they are physically grouped by circuits. However, in the passive IRS cases \cite{MaoZ}-\cite{YanY}, the elements are grouped in the control and operation plane, i.e., the elements in the same group have the same phase-shift coefficient.

Motivated by the above discussion, in this paper, we focus on the hardware-and-energy efficient sub-connected active RIS architecture presented in \cite{KLiu}, where multiple reconfigurable electromagnetic elements are grouped into a sub-array connected to one single amplifier. Particularly, we first propose a reformative and accurate signal model for the sub-connected active RIS, which describes the flow of signal more reasonably and realistically. Afterwards, we investigate the application of sub-connected active RIS in multiuser multi-input single-output (MU-MISO) wireless communication systems for achieving higher sum-rate and less power consumption. The main contributions of this paper can be summarized as follows:
\begin{itemize}
  \item We derive a more accurate and practical signal model for the sub-connected active RIS by considering the combination and re-distribution of the incident signals amplified by the same amplifier.
  \item Then, we employ the sub-connected active RIS in an MU-MISO system and investigate the sum-rate maximization problem, whose goal is to maximize the sum-rate of the system under the consideration of the transmit power constraint at the base station (BS) and total consumed power constraint at the active RIS. An effective joint design algorithm is proposed to solve for the transmit precoding and the RIS reflection beamforming (i.e., the reflection phase-shift and amplification coefficients) by utilizing fractional programming (FP), block coordinate descent (BCD), alternating direction method of multipliers (ADMM), and majorization-minimization (MM) methods.
  \item The power minimization problem is also studied to minimize the total power consumption of BS and active RIS subject to the signal-to-interference-plus-noise ratio (SINR) requirements of all users. In order to handle the resulting non-convex problem, we transform the original problem into several more tractable sub-problems, which can be treated by BCD framework. After some sophisticated matrix manipulations, the second-order cone programming (SOCP) method and ADMM-MM based algorithm are developed to alteratively solve the sub-problems.
  \item Finally, extensive simulation results verify the effectiveness of our proposed algorithms and validate the advantages of the sub-connected active RIS over the traditional fully-connected one. To be specific, the sub-connected active RIS requires only about 3\% active components as the fully-connected structure to achieve similar sum-rate performance. Moreover, compared with the fully-connected active RIS, the sub-connected scheme with the proposed algorithm requires only 85\% power consumption and 1/8 amplifiers to provide the same level of SINR performance of each user.
\end{itemize}

\textit{Notations}: $a$ is a scalar, $\mathbf{a}$ is a column vector, and $\mathbf{A}$ is a matrix, respectively. $\mathbf{A}^{T}$, $\mathbf{A}^{*}$ and $\mathbf{A}^{H}$ denote the transpose, conjugate, and Hermitian (conjugate transpose) operations, respectively. $\mathbb{R}$ and $\mathbb{C}$ represent the sets of real numbers and complex numbers, respectively. $|a|$, $\|\mathbf{a}\|$ and $\|\mathbf{A}\|_F$ denote the magnitude of a scalar $a$, the norm of a vector $\mathbf{a}$ and the Frobenius norm of matrix $\mathbf{A}$. $\operatorname{diag}(\mathbf{a})$ is a diagonal matrix whose diagonal elements are extracted from vector $\mathbf{a}$. $\operatorname{blkdiag}(\mathbf{A}_1, \cdots, \mathbf{A}_N)$ denotes a block diagonal matrix which is composed of matrices $\mathbf{A}_n$, $n = 1, \cdots, N$. $\operatorname{Tr}\{\mathbf{A}\}$ is the trace of the matrix $\mathbf{A}$ and $\operatorname{vec}\{\mathbf{A}\}$ denotes vectorization of the matrix $\mathbf{A}$. $\mathbf{I}_{N}$ is an identity matrix of $N$ dimension and $\mathbf{1}_{N}$ refers to an $N$-dimension all-ones vector. $\otimes$ is the Kronecker product. $\Re\{\cdot\}$ and $\Im\{\cdot\}$ extract the real part and imaginary part of a complex number, respectively.

\section{Sub-Connected Active RIS and System Model}
\subsection{Architecture of Sub-Connected Active RIS}

\begin{figure}[t]
\subfigure[Active RIS with fully-connected array architecture.]{
\begin{minipage}[t]{1\linewidth}
\includegraphics[width=2.9 in]{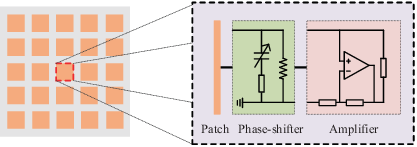}
\centering
\end{minipage}%
\label{fig:fully}
}%

\subfigure[Active RIS with sub-connected array architecture.]{
\begin{minipage}[t]{1\linewidth}
\includegraphics[width=3.2 in]{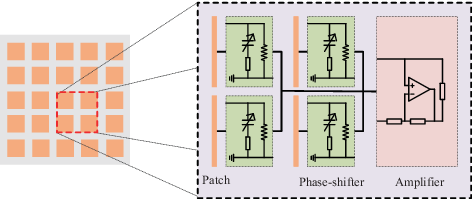}
\centering
\end{minipage}%
\label{fig:sub}
}%
\centering
\caption{Comparison of different active RIS architectures.}
\label{fig:structures}
\vspace{-0.4cm}
\end{figure}

The traditional structure of active RIS is illustrated in Fig. \ref{fig:fully}, where each reconfigurable electromagnetic element (equivalently depicted as a patch and a phase-shifter) integrates one dedicated reflection-type amplifier to amplify the incident signal. However, this fully-connected structure requires plenty of active amplifiers as the number of RIS elements increases, which results in enormous hardware cost and power consumption. For instance, the hardware static power of a typical reflection-type amplifier is 10mW \cite{KLiu}, and the total consumed power of a 512-element active RIS will be up to 5.12W, which is uneconomical for realistic applications. This shortcoming motivates researchers to develop a more cost-effective sub-connected structure, where multiple RIS elements are driven by only one amplifier, as shown in Fig. \ref{fig:sub}. Consequently, the sub-connected array architecture is more hardware and energy efficient than the fully-connected structure owing to the significant reduction in the number of amplifiers. Nevertheless, this sub-connected architecture may cause performance loss since the DoFs of active RIS also decrease, which will be investigated in the rest of this paper.

\subsection{Signal Modeling of Sub-Connected Active RIS}
While the sub-connected structure of active RIS is more hardware efficient, it results in complicated signal modeling, which will be described in details as follows.
Specifically, we divide the $M$ elements of the RIS evenly into $L$ sub-arrays, each of which is connected to a single amplifier. In other words, $Q = M/L$ elements within a sub-array are driven by the same amplifier. Let $\boldsymbol{\theta} \triangleq [\theta_1, \theta_2, \cdots, \theta_M]^T$ with $|\theta_m| = 1,m = 1, \cdots, M$, denote the phase-shift coefficients of the active RIS. Then, we further denote $\tilde{\boldsymbol{\theta}}_l \triangleq [\theta_{(l-1)Q+1}, \cdots, \theta_{lQ}]^T$ as the phase-shift coefficient vector of the $l$-th sub-array.
With the incoming signal $\mathbf{x}_\mathrm{in} \triangleq [x_1, \cdots, x_M]^T$ of all $M$ elements of the active RIS, we denote $\mathbf{x}_{l} \triangleq [x_{(l-1)Q+1}, \cdots, x_{lQ}]^{T}$ as the incident signal of the $l$-th sub-array. The signals of each element of the $l$-th sub-array first pass through their corresponding phase-shifter circuits and then combined as one signal $x_{\mathrm{c},l} = \sum_{i = (l-1)Q+1}^{lQ} x_i e^{j\theta_i} = \tilde{\boldsymbol\theta}_l^T\mathbf{x}_{l}$ for signal amplification using only one amplifier. Afterwards, the amplified compound signal can be expressed as $x_{\mathrm{a},l} = a_lx_{\mathrm{c},l}$, where $a_l \geq 0, l = 1,\cdots, L$, represents the amplification coefficient for the $l$-th sub-array\footnote{We assume that the amplifier operates in its linear region with no limits on the incident signal power in this work, and the investigation on the dependency between the incident signal power and amplification gain is worthy pursuing in the future.}. Furthermore, the signal $x_{\mathrm{a},l}$ is split into $Q$ paths with equally distributed powers. The signal of each path feeds back to each electromagnetic element and is emitted with corresponding phase-shift. This process can be modeled as $\mathbf{y}_l = \frac{1}{\sqrt{Q}} \tilde{\boldsymbol\theta}_l x_{\mathrm{a},l} = \frac{1}{\sqrt{Q}} a_l\tilde{\boldsymbol\theta}_l \tilde{\boldsymbol\theta}_l^T\mathbf{x}_{l}$, where the signal is scaled by $\frac{1}{\sqrt{Q}}$ since the amplified signal is equally split for $Q$ elements of the sub-array\footnote{In this paper, we consider a simple fixed and equal power distribution for hardware efficiency. Actually, reconfigurable power divider can adjust the power allocation dynamically, which will be investigated in our future work.} \cite{Tasci}.
To sum up, the signal model of the $l$-th sub-array is given by $\mathbf{y}_l = \tilde{\mathbf{\Psi}}_l\mathbf{x}_{l}$ with the definition of the beamforming for the $l$-th sub-array as $\tilde{\mathbf{\Psi}}_l \triangleq \frac{1}{\sqrt{Q}} a_l\tilde{\boldsymbol{\theta}}_l\tilde{\boldsymbol{\theta}}_l^T$. The output signal $\mathbf{y}_\mathrm{out}$ (i.e., the reflected signal by the active RIS) is expressed as
\begin{equation}\label{eq:signal_model}
  \mathbf{y}_\mathrm{out} = [\tilde{\mathbf{\Psi}}_1\mathbf{x}_{1}, \cdots, \tilde{\mathbf{\Psi}}_L\mathbf{x}_{L}]^T= \mathbf{\Psi} \mathbf{x}_\mathrm{in},
\end{equation}
in which we define the reflection beamforming matrix of the sub-connected active RIS as $\mathbf{\Psi} \triangleq \operatorname{blkdiag}{(\tilde{\mathbf{\Psi}}_1, \cdots, \tilde{\mathbf{\Psi}}_L)}$. In order to present $\mathbf{\Psi}$ in an explicit form with the phase-shift coefficients and amplification factors, we construct the combined reflection phase-shift matrix as $\tilde{\mathbf{\Theta}} \triangleq \operatorname{blkdiag} (\tilde{\boldsymbol{\theta}}^T_1, \cdots, \tilde{\boldsymbol{\theta}}^T_L )$ and denote $\mathbf{a} \triangleq [a_1, a_2, \cdots, a_L]^T$ and $\mathbf{A} \triangleq  \operatorname{diag}{(\mathbf{a})}$ as the amplification coefficients vector and  the amplification matrix of the RIS. Then, the reflection beamforming of the sub-connected active RIS can be re-formulated as
\begin{equation}\label{eq:Psi}
  \mathbf{\Psi} \triangleq \frac{1}{\sqrt{Q}}\tilde{\mathbf{\Theta}}^T\mathbf{A}\tilde{\mathbf{\Theta}}.
\end{equation}
Unlike the simple and rough signal model presented in \cite{KLiu}, which loosely assumes that the incoming signals on the same sub-array can be amplified independently without interfering each other, our proposed signal model precisely describes the realistic signals combination and re-distribution process due to the use of only one amplifier. More importantly, we should emphasize that the double phase-shift $\tilde{\mathbf{\Theta}}$ in $\mathbf{\Psi}$ introduces significant difficulty in the joint design of phase-shift $\tilde{\mathbf{\Theta}}$ and amplification $\mathbf{A}$ to generate optimal reflection beamforming $\mathbf{\Psi}$ of the sub-connected active RIS.

\subsection{System Model}
\begin{figure}[t]
\centering
  \includegraphics[width = 3.2 in]{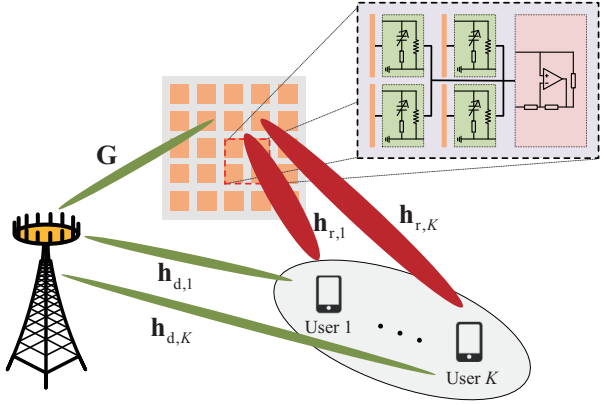}
  \caption{A sub-connected active RIS-aided MU-MISO system.}
  \vspace{-0.1cm}
  \label{fig:system model}
  \vspace{-0.2cm}
\end{figure}
We consider an RIS-aided MU-MISO system as shown in Fig. \ref{fig:system model}, where an $N$-antenna BS communicates $K$ single-antenna users with the aid of an $M$-element active sub-connected RIS, $N \geq K$.
Denote $\mathbf{h}_{\mathrm{d}, k}$, $\mathbf{G}$ and $\mathbf{h}_{\mathrm{r}, k}$ as the channels from the BS to the user-$k$, from the BS to the RIS, and from the RIS to the user-$k$, respectively. We assume that the instantaneous channel state information (CSI) for all channels is perfectly available to the BS. In practice, the CSI can be estimated by various existing efficient channel estimation approaches \cite{HuC}, \cite{ZhengB}-\cite{WangZ}. Then, the received signal at the user-$k$ can be expressed as
\vspace{-0.2cm}
\begin{equation}\label{eq:received_signal_at_user_k}
  y_k = (\mathbf{h}_{\mathrm{d}, k}^H + \mathbf{h}_{\mathrm{r}, k}^H\mathbf{\Psi}\mathbf{G})\sum_{i=1}^{K}\mathbf{w}_is_i + \mathbf{h}_{\mathrm{r}, k}^H\mathbf{\Psi}\mathbf{z} + n_k,
\end{equation}
where $s_i$ is the transmit symbol for the user-$i$. It is assumed that $s_i, i = 1, \cdots, K$, are independent symbols with zero mean and unit variance, i.e., $\mathbb{E}\{s_is_i^{*}\} =1$ and $\mathbb{E}\{s_is_j^{*}\} =0, ~ \forall i \neq j$. $\mathbf{w}_i \in \mathbb{C}^{N}$ is corresponding precoding vector for the user-$i$. $\mathbf{z} \sim \mathcal{C} \mathcal{N} (\mathbf{0}, \sigma^{2}_z\mathbf{I}_{M})$ and $n_k \sim \mathcal{C} \mathcal{N} (0, \sigma_k^{2})$ are the dynamic noise at the active RIS and the additive white Gaussian noise (AWGN) at the user-$k$, respectively.  Accordingly, the SINR at the $k$-th user can be written as
\begin{equation}\label{eq:SINR_at_user_k}
  \gamma_{k}=\frac{|\mathbf{h}_{k}^{H} \mathbf{w}_{k}|^{2}}{\sum_{i=1, i \neq k}^{K}|\mathbf{h}_{k}^{H} \mathbf{w}_{i}|^{2}+\|\mathbf{h}_{\mathrm{r}, k}^{H} \mathbf{\Psi}\|^{2} \sigma_{z}^{2}+\sigma_k^{2}},
\end{equation}
where $\mathbf{h}_{k}^{H} \triangleq \mathbf{h}_{\mathrm{d}, k}^{H} + \mathbf{h}_{\mathrm{r}, k}^{H} \mathbf{\Psi} \mathbf{G}$ represents the composite channel from the BS to the user-$k$.
\subsection{Power Model}
The total power consumption at the BS comprises the transmit power as well as the hardware static power, which is formulated as
\begin{equation}\label{eq:Pbs}
  \mathcal{P}_\mathrm{b} = \nu_1^{-1}\sum_{k=1}^{K}\|\mathbf{w}_{k}\|^{2} + W_\mathrm{BS},
\end{equation}
where $\nu_1$ accounts for the energy conversion efficiency of the devices, $W_\mathrm{BS}$ denotes the dissipated power consumed at the BS. Similarly, the power consumption at the sub-connected active RIS can be expressed as
\begin{equation}\label{eq:Pris}
  \mathcal{P}_\mathrm{r} =  \nu_2^{-1}\Big(\sum_{k=1}^{K}\|\mathbf{\Psi} \mathbf{G} \mathbf{w}_{k}\|^{2}+\|\mathbf{\Psi}\|^{2}_F \sigma_{z}^{2}\Big) + M W_\mathrm{PS} + L W_\mathrm{PA},
\end{equation}
where $\nu_2$ has the same definition as $\nu_1$, $W_\mathrm{PS}$ and $W_\mathrm{PA}$ represent the static powers consumed by the phase-shift circuit and the amplifier at the active RIS, respectively.

\section{Joint Design for Sum-Rate Maximization Problem}
\subsection{Problem Formulation}
In this section, we aim to maximize the sum-rate of the MU-MISO system by jointly designing transmit precoding vectors $\mathbf{w}_k$, reflection phase-shift coefficients $\boldsymbol{\theta}$ and reflection amplification factors $\mathbf{a}$, while satisfying the maximum power consumption constraints at the BS and the active RIS. Thus, the optimization problem is formulated as
\begin{subequations}\label{pr:original_problem}
\begin{align}
\max_{\mathbf{w}_k, \boldsymbol{\theta}, \mathbf{a}} ~~&\sum\nolimits_{k=1}^{K} \log _{2}(1+{\gamma}_{k})\\
\text{s.t.} ~~~~& \sum\nolimits_{k=1}^{K}\|\mathbf{w}_{k}\|^{2} \leq P_{\mathrm{BS}},\\
& \nu_2^{-1}\Big(\sum\nolimits_{k=1}^{K}\|\mathbf{\Psi} \mathbf{G} \mathbf{w}_{k}\|^{2}+\|\mathbf{\Psi}\|^{2}_F \sigma_{z}^{2}\Big) \notag\\
&~~~~+ M W_\mathrm{PS} + L W_\mathrm{PA} \leq P_{\mathrm{RIS}}^{\mathrm{tot}},\\
& |\theta_m| = 1,~ \forall m, \\
& a_l \geq 0,~ \forall l,
\end{align}
\end{subequations}
where $P_{\mathrm{BS}}$ represents the maximum available transmit power at the BS and $P_{\mathrm{RIS}}^{\mathrm{tot}}$ represents the maximum available total power at the active RIS, correspondingly, (\ref{pr:original_problem}d) is constant modulus constraint due to phase-shifters, (\ref{pr:original_problem}e) guarantees the amplification coefficients of the amplifiers are nonnegative.

We observe that problem (\ref{pr:original_problem}) is a complicated non-convex problem due to the multi-variable coupling non-convex objective (\ref{pr:original_problem}a) and constraint (\ref{pr:original_problem}c), and non-convex constant-modulus constraint (\ref{pr:original_problem}d). To tackle these difficulties, we propose to equivalently reformulate the original problem through the theory of FP, and then divide the transformed problem into several more tractable sub-problems that can be optimized alternatively via the BCD method.

\subsection{Fractional Programming Transform}
In this subsection, we attempt to deal with the non-convex objective function in (\ref{pr:original_problem}) via FP algorithm. Specifically, based on the \textit{Lagrangian Dual Transform} \cite{ShenK}, the objective in (\ref{pr:original_problem}) is equivalent to
\vspace{-0.2cm}
\begin{equation}\label{eq:FP}
\begin{aligned}
  &f_1(\mathbf{w}_k, \boldsymbol{\theta}, \mathbf{a}, \boldsymbol{\mu}) \triangleq \sum_{k=1}^{K} \log_2(1+\mu_k) - \sum_{k=1}^{K} \mu_k\\
   &~~~~~~~~+ \sum_{k=1}^{K}\frac{(1+\mu_k)|\mathbf{h}_{k}^{H} \mathbf{w}_{k}|^{2}}{\sum_{i=1}^{K}|\mathbf{h}_{k}^{H} \mathbf{w}_{i}|^{2}+\|\mathbf{h}_{\mathrm{r}, k}^{H} \mathbf{\Psi}\|^{2} \sigma_{z}^{2}+\sigma_k^{2}},
\end{aligned}
\end{equation}
when the auxiliary variable $\boldsymbol{\mu} \triangleq [\mu_1, \mu_2, \cdots, \mu_K]^T$ has the optimal solution as
\begin{equation}\label{eq:optimal_mu}
  \mu_k^{\star} =\frac{|\mathbf{h}_{k}^{H} \mathbf{w}_{k}|^{2}}{\sum_{i=1, i \neq k}^{K}|\mathbf{h}_{k}^{H} \mathbf{w}_{i}|^{2}+\|\mathbf{h}_{\mathrm{r}, k}^{H} \mathbf{\Psi}\|^{2} \sigma_{z}^{2}+\sigma_k^{2}}, ~\forall k.
\end{equation}
Unfortunately, since the last term of (\ref{eq:FP}) is a sum of multiple fractions, the transformed problem is still challenging to deal with directly. As a consequence, in order to solve it in a more effective way, we apply \textit{Quadratic Transform} \cite{ShenK} on the fractional term and further equivalently formulate the objective function (\ref{eq:FP}) as
\begin{equation}\label{eq:FP2}
\begin{aligned}
  &f_2(\mathbf{w}_k, \boldsymbol{\theta}, \mathbf{a}, \boldsymbol{\mu}, \boldsymbol{\eta}) \\
  &~\triangleq \sum\nolimits_{k=1}^{K}\Big[\log_2 \left(1+\mu_{k}\right)-\mu_{k}+2 \sqrt{1+\mu_{k}} \Re\left\{\eta_{k}^{*} \mathbf{h}_{k}^{H} \mathbf{w}_{k}\right\} \\ &~~~~ -\left|\eta_{k}\right|^{2}\Big(\sum\nolimits_{i=1}^{K}\left|\mathbf{h}_{k}^{H} \mathbf{w}_{i}\right|^{2}+\left\|\mathbf{h}_{\mathrm{r}, k}^{H} \mathbf{\Psi}\right\|^{2} \sigma_{z}^{2}+\sigma_k^{2}\Big)\Big],
\end{aligned}
\end{equation}
where the auxiliary variable $\boldsymbol{\eta} \triangleq [\eta_1, \eta_2, \cdots, \eta_K]^T$ has the following optimal value
\begin{equation}\label{eq:optimal_eta}
  \eta_{k}^{\star}=\frac{\sqrt{1+\mu_{k}} \mathbf{h}_{k}^{H} \mathbf{w}_{k}}{\sum_{i=1}^{K}\left|\mathbf{h}_{k}^{H} \mathbf{w}_{i}\right|^{2}+\left\|\mathbf{h}_{\mathrm{r}, k}^{H} \mathbf{\Psi}\right\|^{2} \sigma_{z}^{2}+\sigma_k^{2}},~\forall k.
\end{equation}

Through the above analysis, the optimization problem (\ref{pr:original_problem}) is re-written as
\begin{equation}\label{pr:fianl_problem}
\begin{aligned}
\max_{\mathbf{w}_k, \boldsymbol{\theta}, \mathbf{a}, \boldsymbol{\mu}, \boldsymbol{\eta}} ~~&f_2(\mathbf{w}_k, \boldsymbol{\theta}, \mathbf{a}, \boldsymbol{\mu}, \boldsymbol{\eta})\\
\text{s.t.} ~~~~~~& \sum\nolimits_{k=1}^{K}\|\mathbf{w}_{k}\|^{2}\leq {P}_{\mathrm{BS}},\\
& \sum\nolimits_{k=1}^{K}\|\mathbf{\Psi} \mathbf{G} \mathbf{w}_{k}\|^{2}+\|\mathbf{\Psi}\|^{2}_F \sigma_{z}^{2} \leq P_{\mathrm{RIS}},\\
& |\theta_m| = 1,~ \forall m, \\
& a_l \geq 0,~ \forall l,
\end{aligned}
\end{equation}
where
we define $P_{\mathrm{RIS}} \triangleq \nu_2(P_{\mathrm{RIS}}^{\mathrm{tot}} - M W_\mathrm{PS} - L W_\mathrm{PA})$ for brevity. To effectively solve problem (\ref{pr:fianl_problem}), we propose to adopt the BCD method to alternately update $\boldsymbol{\mu}$, $\boldsymbol{\eta}$, $\mathbf{w}_k$, $\boldsymbol{\theta}$ and $\mathbf{a}$ with fixed others, which is presented in details as follows.

\subsection{Block update}
\textit{1) Update auxiliary variables $\boldsymbol{\mu}$ and $\boldsymbol{\eta}$:}
Fixing $\mathbf{w}_k$, $\boldsymbol{\theta}$ and $\mathbf{a}$, the auxiliary variables $\boldsymbol{\mu}$ and $\boldsymbol{\eta}$ should be updated as in (\ref{eq:optimal_mu}) and (\ref{eq:optimal_eta}), respectively.

\textit{2) Update precoding $\mathbf{w}_k$:} With given $\boldsymbol{\mu}$, $\boldsymbol{\eta}$, $\boldsymbol{\theta}$ and $\mathbf{a}$, the optimization problem with respect to $\mathbf{w}_k$ can be simplified to
\begin{equation}\label{pr:w}
\begin{aligned}
\max_{\mathbf{w}_k} &~~~~f_2(\mathbf{w}_k)\\
\text{s.t.} ~&~~ \sum\nolimits_{k=1}^{K}\|\mathbf{w}_{k}\|^{2} \leq {P}_{\mathrm{BS}},\\
& ~~\sum\nolimits_{k=1}^{K}\|\mathbf{\Psi} \mathbf{G} \mathbf{w}_{k}\|^{2} +\|\mathbf{\Psi}\|^{2}_F \sigma_{z}^{2} \leq P_{\mathrm{RIS}}.
\end{aligned}
\end{equation}
For conciseness, we define $\mathbf{w} \triangleq  [\mathbf{w}_1^T,\mathbf{w}_2^T, \cdots, \mathbf{w}_K^T]^T$. By dropping the constant terms independent of $\mathbf{w}$, problem (\ref{pr:w}) can be re-formulated as
\begin{equation}\label{pr:final_w}
\begin{aligned}
\max_{\mathbf{w}} ~~~&\Re\{\mathbf{x}^H\mathbf{w}\} - \mathbf{w}^H\mathbf{Y}\mathbf{w}\\
\text{s.t.} ~~~~& \mathbf{w}^H\mathbf{w} \leq {P}_{\mathrm{BS}},\\
& \mathbf{w}^H\mathbf{Z}\mathbf{w} \leq P_{\mathrm{RIS}} - \|\mathbf{\Psi}\|^{2}_F \sigma_{z}^{2},
\end{aligned}
\end{equation}
where for brevity we define
\begin{subequations}\label{eq:w1}
\begin{align}
\mathbf{x} &\triangleq  [\mathbf{x}_1^T,\mathbf{x}_2^T, \cdots, \mathbf{x}_K^T]^T,~\mathbf{x}_k \triangleq 2\sqrt{1+\mu_{k}}\eta_{k}\mathbf{h}_{k},\\
\mathbf{Y} &\triangleq \mathbf{I}_{K} \otimes \left( \sum\nolimits_{k = 1}^{K}|\eta_k|^2\mathbf{h}_k\mathbf{h}_k^H\right),\\
\mathbf{Z} &\triangleq \mathbf{I}_{K} \otimes\left(\mathbf{G}^{H} \mathbf{\Psi}^{H} \mathbf{\Psi} \mathbf{G}\right).
\end{align}
\end{subequations}
It is obvious that problem (\ref{pr:final_w}) is a standard quadratic constraint quadratic programming (QCQP) problem, and thus the optimal $\mathbf{w}$ can be easily obtained by various existing algorithms or toolboxes like CVX \cite{Grant}.

\textit{3) Update RIS reflection phase-shift coefficients $\boldsymbol{\theta}$:}
After obtaining $\boldsymbol{\mu}$, $\boldsymbol{\eta}$, $\mathbf{w}_k$ and $\mathbf{a}$, the problem of optimizing the reflection phase-shift coefficients $\boldsymbol{\theta}$ is reduced to
\begin{subequations}\label{pr:Theta}
\begin{align}
\max_{\boldsymbol{\theta}} ~~&~~f_2(\boldsymbol{\theta})\\
\text{s.t.} ~~~& \sum\nolimits_{k=1}^{K}\|\mathbf{\Psi} \mathbf{G} \mathbf{w}_{k}\|^{2} + \|\mathbf{\Psi}\|^{2}_F \sigma_{z}^{2} \leq P_{\mathrm{RIS}},\\
& ~|\theta_m| = 1,~ \forall m.
\end{align}
\end{subequations}
Obviously, problem (\ref{pr:Theta}) cannot be directly solved due to the implicit function (\ref{pr:Theta}a) relevant to $\boldsymbol{\theta}$ and the non-convex unit modulus constraint (\ref{pr:Theta}c).

In order to handle above issues, we first investigate the ADMM method \cite{Boyd} to solve for $\boldsymbol{\theta}$ under the unit modulus constraint. Specifically, we introduce an auxiliary variable $\boldsymbol{\vartheta} \triangleq [\vartheta_1, \cdots, \vartheta_M]^T \in \mathbb{C}^{M}$ to transform the problem (\ref{pr:Theta}) into
\begin{subequations}\label{pr:Theta_ADMM}
\begin{align}
\max_{\boldsymbol{\theta}, \boldsymbol{\vartheta}} ~~&~~f_2(\boldsymbol{\theta})\\
\text{s.t.} ~~~& \sum\nolimits_{k=1}^{K}\|\mathbf{\Psi} \mathbf{G} \mathbf{w}_{k}\|^{2} + \|\mathbf{\Psi}\|^{2}_F \sigma_{z}^{2} \leq P_{\mathrm{RIS}},\\
& ~|\theta_m| \leq 1,~ \forall m,\\
& ~\boldsymbol{\theta} = \boldsymbol{\vartheta},\\
& ~|\vartheta_m| = 1,~ \forall m,
\end{align}
\end{subequations}
whose solution can be obtained by solving its augmented Lagrangian problem:
\begin{subequations}\label{pr:Theta_ADMM_AL}
\begin{align}
\max_{\boldsymbol{\theta}, \boldsymbol{\vartheta}, \boldsymbol{\omega}} ~~&~~f_2(\boldsymbol{\theta}) - \frac{\rho}{2}\|\boldsymbol{\theta} - \boldsymbol{\vartheta} + \frac{\boldsymbol{\omega}}{\rho}\|^2\\
\text{s.t.} ~~~& \sum\nolimits_{k=1}^{K}\|\mathbf{\Psi} \mathbf{G} \mathbf{w}_{k}\|^{2} + \|\mathbf{\Psi}\|^{2}_F \sigma_{z}^{2} \leq { P}_{\mathrm{RIS}},\\
& ~|\theta_m| \leq 1,~ \forall m,\\
& ~|\vartheta_m| = 1,~ \forall m,
\end{align}
\end{subequations}
where $\rho > 0$ is a penalty parameter and $\boldsymbol{\omega} \in \mathbb{C}^M$ is the dual variable. Compared to problem (\ref{pr:Theta_ADMM}), problem (\ref{pr:Theta_ADMM_AL}) is more tractable after removing the equality constraint (\ref{pr:Theta_ADMM}d).
For this multi-variable problem, we propose to alternately update each variable given the others to solve it efficiently.

\textbf{Update $\boldsymbol{\theta}$:} Given $\boldsymbol{\vartheta}$ and $\boldsymbol{\omega}$, we first re-formulate the problem (\ref{pr:Theta_ADMM_AL}) into an explicit form with respect to $\boldsymbol{\theta}$. Specifically, we first re-written the reflection beamforming matrix as $\mathbf{\Psi} = \frac{1}{\sqrt{Q}}\mathbf{\Theta}\mathbf{E}^T\mathbf{A}\mathbf{E}\mathbf{\Theta}$, where $\mathbf{\Theta} \eqdef \operatorname{diag} (\boldsymbol{\theta})$ is the phase-shift matrix of the RIS and $\mathbf{E} \triangleq \mathbf{I}_{L} \otimes \mathbf{1}_{Q}^T$ is expressed as an indicator matrix accounting for the connection relationship between the reflection elements and the amplifiers. By defining $\mathbf{\Xi} \triangleq \frac{1}{\sqrt{Q}}\mathbf{E}^T\mathbf{A}\mathbf{E}$, $\mathbf{\Psi}$ can be re-expressed as $\mathbf{\Psi} = \mathbf{\Theta}\mathbf{\Xi}\mathbf{\Theta}$. Since the RIS reflection phase-shift matrix $\mathbf{\Theta}$ is diagonal, we have $\mathbf{h}_{\mathrm{r}, k}^{H} \mathbf{\Theta} = \boldsymbol{\theta}^T \operatorname{diag}{(\mathbf{h}_{\mathrm{r}, k}^{*})}$ and $\mathbf{\Theta} \mathbf{G} \mathbf{w}_{k} = \operatorname{diag}{(\mathbf{g}_{k} )}\boldsymbol{\theta}$, where $\mathbf{g}_k \triangleq \mathbf{G}\mathbf{w}_k$. Thus, we can successfully extract the variable $\boldsymbol{\theta}$ from problem (\ref{pr:Theta_ADMM_AL}) and recast it into an explicit problem as:
\begin{subequations}\label{pr:Theta_ADMM_AL_explicit}
\begin{align}
\min_{\boldsymbol{\theta}} ~&~\mathbf{v}^H\mathbf{F}\mathbf{v} + \Re\{\boldsymbol{\theta}^H \mathbf{P}  \boldsymbol{\theta}^{*}\} + \boldsymbol{\theta}^H \mathbf{Q}_1\boldsymbol{\theta} + \frac{\rho}{2}\|\boldsymbol{\theta} - \boldsymbol{\vartheta} + \frac{\boldsymbol{\omega}}{\rho}\|^2\\
\text{s.t.} ~~&~ \boldsymbol{\theta}^H \mathbf{Q}_2\boldsymbol{\theta}  \leq \tau,\\
& ~|\theta_m| \leq 1,~ \forall m,
\end{align}
\end{subequations}
where for brevity we define
\begin{subequations}\label{eq:theta_trans4}
\allowdisplaybreaks[4]
\begin{align}
\mathbf{v} &\triangleq \mathrm{vec}\{\boldsymbol{\theta}\boldsymbol{\theta}^T\} = \boldsymbol{\theta} \otimes \boldsymbol{\theta}, \\
\mathbf{F} &\triangleq \sum_{k = 1}^K |\eta_{k}|^2 \sum_{i = 1}^K \mathbf{f}_{k,i}\mathbf{f}_{k,i}^H,\\
\mathbf{f}_{k,i} &\triangleq \mathrm{vec}\big\{(\operatorname{diag}(\mathbf{h}_{\mathrm{r}, k}^*) \mathbf{\Xi} \operatorname{diag}(\mathbf{g}_{i}))^*\big\}, ~ \forall k,~ \forall i,\\
\mathbf{P} &\triangleq \mathbf{P}_2^H - \mathbf{P}_1^H,\\
\mathbf{P}_1 &\triangleq \sum_{k = 1}^K 2 \sqrt{1+\mu_{k}} \eta_{k}^{*} \operatorname{diag}(\mathbf{h}_{\mathrm{r}, k}^*) \mathbf{\Xi} \operatorname{diag}(\mathbf{g}_{k}),\\
\mathbf{P}_2 &\triangleq \sum_{k = 1}^K |\eta_{k}|^2 \sum_{i = 1}^K 2 (\mathbf{h}_{\mathrm{d},k}^H\mathbf{w}_i)^{*}\operatorname{diag}(\mathbf{h}_{\mathrm{r}, k}^*) \mathbf{\Xi} \operatorname{diag}(\mathbf{g}_{i}),\\
\mathbf{Q}_1 &\triangleq \sum_{k = 1}^K |\eta_{k}|^2 (\operatorname{diag}(\mathbf{h}_{\mathrm{r}, k}^*)\mathbf{\Xi}\mathbf{\Xi}^H \operatorname{diag}(\mathbf{h}_{\mathrm{r}, k}))^{*}\sigma_z^2,\\
\mathbf{Q}_2  &\triangleq \sum\nolimits_{k = 1}^K \operatorname{diag}(\mathbf{g}_{k}^*)\mathbf{\Xi}^H \mathbf{\Xi}\operatorname{diag}(\mathbf{g}_{k}),\\
\tau &\triangleq P_{\mathrm{RIS}} - \|\mathbf{\Psi}\|_F^2 \sigma_{z}^{2}.
\end{align}
\end{subequations}
In particular, $\mathbf{\Theta}$ is a diagonal matrix with unit-modulus diagonal elements and thus we have $\|\mathbf{\Psi}\|^{2}_F = \|\mathbf{\Xi}\|^{2}_F$. Then, $\tau$ can be further expressed as $\tau = {P}_{\mathrm{RIS}} - \|\mathbf{\Xi}\|_F^2 \sigma_{z}^{2}$, which is irrelevant to $\boldsymbol{\theta}$.

We can notice that the objective function (\ref{pr:Theta_ADMM_AL_explicit}a) is still challenging to directly optimize due to the quartic term $\mathbf{v}^H\mathbf{F}\mathbf{v}$ with respect to $\boldsymbol{\theta}$ and real-valued term $\Re\{\boldsymbol{\theta}^H \mathbf{P}  \boldsymbol{\theta}^{*}\}$. In the following, we apply MM algorithm \cite{SunY} to seek for a more tractable surrogate function for these non-convex terms via the \textit{second-order Taylor expansion}.

Specifically, since $\mathbf{F}$ is a positive semidefinite Hermitian matrix as defined in (\ref{eq:theta_trans4}b), we can construct a surrogate function of $\mathbf{v}^H\mathbf{F}\mathbf{v}$ at point $\mathbf{v}_t$ (i.e., solution $\boldsymbol{\theta}_t$ in the $t$-th iteration) as
\begin{equation}\label{eq:MM}
\begin{aligned}
  \mathbf{v}^H\mathbf{F}\mathbf{v} &\leq \lambda_\mathrm{f} \mathbf{v}^H\mathbf{v} + 2\Re\{\mathbf{v}^H(\mathbf{F} - \lambda_\mathrm{f}\mathbf{I}_{M^2})\mathbf{v}_{t}\} \\
  &\quad+ \mathbf{v}_t^H(\lambda_\mathrm{f}\mathbf{I}_{M^2} - \mathbf{F})\mathbf{v}_{t},
\end{aligned}
\end{equation}
where $\lambda_\mathrm{f}$ is an upper-bound of the eigenvalues of $\mathbf{F}$. Due to the high complexity $\mathcal{O}(M^6)$ of eigenvalue decomposition of matrix $\mathbf{F}$, we choose $\lambda_\mathrm{f} = \mathrm{Tr}\{\mathbf{F}\}$ as an efficient and reasonable solution, which is actually the sum of all eigenvalues of the matrix $\mathbf{F}$. Since the matrix $\mathbf{F}$ contains $K^2$ rank-one matrices as expressed in (\ref{eq:theta_trans4}b), we can directly obtain the trace by
\begin{equation}\label{eq:trace}
  \lambda_\mathrm{f} = \mathrm{Tr}\{\mathbf{F}\} = \sum\nolimits_{k = 1}^{K}|\eta_k|^2\sum\nolimits_{i = 1}^K \|\mathbf{f}_{k,i}\|^2.
\end{equation}
Therefore, after considering that the term $\mathbf{v}^H\mathbf{v}$ is upper-bounded by
\begin{equation}\label{eq:MM_v_theta}
\mathbf{v}^H\mathbf{v} = (\boldsymbol{\theta} \otimes \boldsymbol{\theta})^H(\boldsymbol{\theta} \otimes \boldsymbol{\theta}) = (\boldsymbol{\theta}^H\boldsymbol{\theta})\otimes(\boldsymbol{\theta}^H\boldsymbol{\theta}) \leq M^2,
\end{equation}
an appropriate surrogate function of the quartic term $\mathbf{v}^H\mathbf{F}\mathbf{v}$ can be obtained by
\begin{subequations}\label{eq:MM_final}
\begin{align}
  \mathbf{v}^H\mathbf{F}\mathbf{v} &\leq \Re\{\mathbf{v}^H \tilde{\mathbf{f}}_{{t}}\} + c_{\mathrm{f},t} \\
  &= \Re\{\boldsymbol{\theta}^H\tilde{\mathbf{F}}_{{t}} \boldsymbol{\theta}^{*}\} + c_{\mathrm{f},t},
\end{align}
\end{subequations}
where we define
\begin{equation}\label{eq:ft0}
  \tilde{\mathbf{f}}_{{t}} \triangleq 2(\mathbf{F} - \lambda_\mathrm{f} \mathbf{I}_{M^2})\mathbf{v}_{t},
\end{equation}
and $c_{\mathrm{f},t}  \triangleq \lambda_\mathrm{f}  M^2 + \mathbf{v}_t^H(\lambda_\mathrm{f} \mathbf{I}_{M^2} - \mathbf{F})\mathbf{v}_{t} $ is a constant term independent of $\mathbf{v}$ (i.e., $\boldsymbol{\theta}$). In addition, equality (\ref{eq:MM_final}b) holds since we express $\tilde{\mathbf{F}}_{{t}} \in \mathbb{C}^{M \times M}$ in a reshaped version  $\tilde{\mathbf{f}}_{{t}}$, i.e., $\tilde{\mathbf{f}}_{{t}} = \mathrm{vec}\{\tilde{\mathbf{F}}_{{t}}\}$ and utilize $\mathbf{v} \triangleq \boldsymbol{\theta} \otimes \boldsymbol{\theta}$.
Considering the significantly high cost of computing and storing $\tilde{\mathbf{f}}_{{t}}$ which contains an $M^2 \times M^2$-dimensional matrix $\mathbf{F}$, we attempt to directly construct $\tilde{\mathbf{F}}_{{t}}$ using $M \times M$ lower-dimensional matrices. In specific, by utilizing the definitions of $\mathbf{v}$ and $\mathbf{F}$ in (\ref{eq:theta_trans4}a) and (\ref{eq:theta_trans4}b), we can re-express $\tilde{\mathbf{f}}_{{t}}$ as
\begin{subequations}\label{eq:Ft_trans}
\begin{align}
 \tilde{\mathbf{f}}_{{t}} & = 2 \sum_{k = 1}^K |\eta_{k}|^2 \sum_{i = 1}^K \mathrm{vec}\{\tilde{\mathbf{P}}_{k,i}\}\mathrm{vec}^H\{\tilde{\mathbf{P}}_{k,i}\}\mathrm{vec}\{\boldsymbol{\theta}_{t}\boldsymbol{\theta}_{t}^T\} \notag\\
 &\quad ~ -  2\lambda_\mathrm{f} \mathrm{vec}\{\boldsymbol{\theta}_{t}\boldsymbol{\theta}_{t}^T\}\\
 & = 2\sum_{k = 1}^K |\eta_{k}|^2 \sum_{i = 1}^K\mathrm{vec}\{\tilde{\mathbf{P}}_{k,i}\}\mathrm{Tr}\{\boldsymbol{\theta}_{t}\boldsymbol{\theta}_{t}^T\tilde{\mathbf{P}}_{k,i}^H\} \notag \\ &\quad ~  - 2\lambda_\mathrm{f} \mathrm{vec}\{\boldsymbol{\theta}_{t}\boldsymbol{\theta}_{t}^T\}\\
  & = 2\sum_{k = 1}^K |\eta_{k}|^2 \sum_{i = 1}^K \boldsymbol{\theta}_{t}^T\tilde{\mathbf{P}}_{k,i}^H\boldsymbol{\theta}_{t}\mathrm{vec}\{\tilde{\mathbf{P}}_{k,i}\} \notag \\ & \quad ~ - 2\lambda_\mathrm{f} \mathrm{vec}\{\boldsymbol{\theta}_{t}\boldsymbol{\theta}_{t}^T\},
\end{align}
\end{subequations}
where we define $\tilde{\mathbf{P}}_{k,i} \triangleq (\operatorname{diag}(\mathbf{h}_{\mathrm{r}, k}^*) \mathbf{\Xi} \operatorname{diag}(\mathbf{g}_{i}))^{*}$ for conciseness. In addition, (\ref{eq:Ft_trans}b) holds since we use the transformation $\mathrm{Tr}\{\mathbf{A}\mathbf{B}\} = \mathrm{vec}^H\{\mathbf{B}^H\} \mathrm{vec}\{\mathbf{A}\}$. Thus, based on the derivations in (\ref{eq:Ft_trans}), the matrix version of $\tilde{\mathbf{f}}_{{t}}$, i.e., $\tilde{\mathbf{F}}_{{t}}$, can be equivalently written by \begin{equation}\label{eq:Ft}
\tilde{\mathbf{F}}_{{t}} \triangleq 2\sum_{k = 1}^K |\eta_{k}|^2 \sum_{i = 1}^K \boldsymbol{\theta}_{t}^T\tilde{\mathbf{P}}_{k,i}^H\boldsymbol{\theta}_{t}\tilde{\mathbf{P}}_{k,i} - 2\lambda_\mathrm{f} \boldsymbol{\theta}_{t}\boldsymbol{\theta}_{t}^T.
\end{equation}

After obtaining the surrogate function $\Re\{\boldsymbol{\theta}^H\tilde{\mathbf{F}}_{{t}} \boldsymbol{\theta}^{*}\}+ c_{\mathrm{f},t}$ for $\mathbf{v}^H\mathbf{F}\mathbf{v}$, the objective function in (\ref{pr:Theta_ADMM_AL_explicit}) with respect to $\boldsymbol{\theta}$ can be transformed into
\begin{equation}\label{eq:MM_function}
  \min_{\boldsymbol{\theta}}~~ \Re\{\boldsymbol{\theta}^H\mathbf{P}_{{t}}\boldsymbol{\theta}^{*}\}+ \boldsymbol{\theta}^H \mathbf{Q}_1\boldsymbol{\theta} + \frac{\rho}{2}\|\boldsymbol{\theta} - \boldsymbol{\vartheta} + \frac{\boldsymbol{\omega}}{\rho}\|^2
\end{equation}
in which we define $\mathbf{P}_{{t}} \triangleq \tilde{\mathbf{F}}_{{t}} + \mathbf{P}$. In order to handle the non-convex term $\Re\{\boldsymbol{\theta}^H\mathbf{P}_{{t}}\boldsymbol{\theta}^{*}\}$ in (\ref{eq:MM_function}), we propose to convert the complex-valued variable into its real-valued form and apply the second-order Taylor expansion again to seek for a convex surrogate function. Specifically, after defining
\begin{subequations}\label{eq:real_var}
\begin{align}
\bar{\boldsymbol{\theta}} &\triangleq \begin{bmatrix}\Re\{\boldsymbol{\theta}^{T}\} ~~ \Im\{\boldsymbol{\theta}^{T}\}\end{bmatrix}^{T}, \\
\bar{\mathbf{P}}_{{t}} &\triangleq \begin{bmatrix}
\Re\{\mathbf{P}_{{t}}\} & \Im\{\mathbf{P}_{{t}}\} \\
\Im\{\mathbf{P}_{{t}}\} & -\Re\{\mathbf{P}_{{t}}\}
\end{bmatrix},
\end{align}
\end{subequations}
an appropriate surrogate function of $\Re\{\boldsymbol{\theta}^H \mathbf{P}_t  \boldsymbol{\theta}^{*}\}$ can be obtained by
\begin{subequations}\label{eq:MM2}
\begin{align}
  \Re\{\boldsymbol{\theta}^H \mathbf{P}_t  \boldsymbol{\theta}^{*}\} & = \bar{\boldsymbol{\theta}}^T \bar{\mathbf{P}}_{{t}}\bar{\boldsymbol{\theta}},\\
  & \leq \bar{\boldsymbol{\theta}}_t^T\bar{\mathbf{P}}_{{t}}\bar{\boldsymbol{\theta}}_t + \bar{\boldsymbol{\theta}}_t^T(\bar{\mathbf{P}}_{{t}} + \bar{\mathbf{P}}_{{t}}^T)(\bar{\boldsymbol{\theta}} - \bar{\boldsymbol{\theta}}_t) \notag\\
  & \quad + \frac{\lambda_{\mathrm{p},t}}{2}(\bar{\boldsymbol{\theta}} - \bar{\boldsymbol{\theta}}_t)^T(\bar{\boldsymbol{\theta}} - \bar{\boldsymbol{\theta}}_t),\\
  & = \frac{\lambda_{\mathrm{p},t}}{2}\bar{\boldsymbol{\theta}}^T\bar{\boldsymbol{\theta}} + \Re\{\bar{\boldsymbol{\theta}}^T \bar{\mathbf{p}}_t\} + c_{\mathrm{p},t},\\
  & = \frac{\lambda_{\mathrm{p},t}}{2}\boldsymbol{\theta}^H\boldsymbol{\theta} + \Re\{\boldsymbol{\theta}^H \mathbf{U}\bar{\mathbf{p}}_t\}+ c_{\mathrm{p},t},
\end{align}
\end{subequations}
where $\lambda_{\mathrm{p},t}$ is the maximum eigenvalue of the Hessian matrix $(\bar{\mathbf{P}}_{{t}} + \bar{\mathbf{P}}_{{t}}^T)$, $\bar{\mathbf{p}}_t \triangleq (\bar{\mathbf{P}}_{{t}} + \bar{\mathbf{P}}_{{t}}^T - \lambda_{\mathrm{p},t}\mathbf{I}_{2M})\bar{\boldsymbol{\theta}}_t$, $\mathbf{U} \triangleq \left[\mathbf{I}_{M} ~ \jmath\mathbf{I}_{M}\right]$ and $c_{\mathrm{p},t}$ is a constant independent of $\bar{\boldsymbol{\theta}}$.

Given the surrogate function derived for $\Re\{\boldsymbol{\theta}^H \mathbf{P}_t  \boldsymbol{\theta}^{*}\}$ in (\ref{eq:MM2}d), a convex upper-bound function of (\ref{eq:MM_function}) can be expressed as
\begin{equation}\label{eq:theta_final_MM+MM}
\begin{aligned}
  g(\boldsymbol{\theta}|\boldsymbol{\theta}_t)& = \frac{\lambda_{\mathrm{p},t}}{2}\boldsymbol{\theta}^H\boldsymbol{\theta} + \Re\{\boldsymbol{\theta}^H \mathbf{U}\bar{\mathbf{p}}_t\} + \boldsymbol{\theta}^H \mathbf{Q}_1\boldsymbol{\theta}\\
   & \quad~+ \frac{\rho}{2}\|\boldsymbol{\theta} - \boldsymbol{\vartheta} + \frac{\boldsymbol{\omega}}{\rho}\|^2 + c_{\mathrm{p},t},\\
  & = \boldsymbol{\theta}^H \mathbf{\Upsilon}_t \boldsymbol{\theta} + \Re\{\boldsymbol{\theta}^H \boldsymbol{\zeta}_t\} + c_{\mathrm{s},t},
\end{aligned}
\end{equation}
where for ease of notation, we define $\mathbf{\Upsilon}_t \triangleq \mathbf{Q}_1 + \frac{\lambda_{\mathrm{p},t} + \rho}{2}\mathbf{I}_M$, $\boldsymbol{\zeta}_t \triangleq \mathbf{U}\bar{\mathbf{p}}_t - \rho \boldsymbol{\vartheta} + \boldsymbol{\omega}$, and $c_{\mathrm{s},t} \triangleq c_{\mathrm{p},t} + \frac{\rho}{2}\|\boldsymbol{\vartheta} - \boldsymbol{\omega}/\rho\|^2$. Thus, the RIS reflection phase-shift coefficients design problem at the $(t+1)$-th iteration can be re-formulated as
\begin{subequations}\label{pr:theta_final}
\begin{align}
  \min_{\boldsymbol{\theta}} & ~~~ \boldsymbol{\theta}^H \mathbf{\Upsilon}_t \boldsymbol{\theta} + \Re\{\boldsymbol{\theta}^H\boldsymbol{\zeta}_t\} \\
  \text{s.t.} &~~~ \boldsymbol{\theta}^H \mathbf{Q}_2\boldsymbol{\theta}  \leq \tau,\\
        & ~~~|\theta_m| \leq 1,~ \forall m,
\end{align}
\end{subequations} which can be optimally solved by existing convex optimization solvers.

\textbf{Update $\boldsymbol{\vartheta}$:} Fixing $\boldsymbol{\theta}$ and $\boldsymbol{\omega}$, the optimization problem of solving for the auxiliary variable $\boldsymbol{\vartheta}$ is
\begin{equation}\label{pr_vartheta}
\begin{aligned}
\min_{\boldsymbol{\vartheta}} & ~~~ \frac{\rho}{2}\|\boldsymbol{\theta} - \boldsymbol{\vartheta} + \frac{\boldsymbol{\omega}}{\rho}\|^2\\
\text{s.t.} &~~~ |\vartheta_m| = 1,~ \forall m.
\end{aligned}
\end{equation}
Thus, the optimal $\boldsymbol{\vartheta}$ should be updated by the phase alignment
\begin{equation}\label{eq:vartheta}
  \boldsymbol{\vartheta} = e^{\jmath \angle(\rho\boldsymbol{\theta} + \boldsymbol{\omega})}.
\end{equation}

\textbf{Update $\boldsymbol{\omega}$:} With obtained $\boldsymbol{\theta}$ and $\boldsymbol{\vartheta}$, the dual variable $\boldsymbol{\omega}$ can be updated by
\begin{equation}\label{eq:omega}
  \boldsymbol{\omega} := \boldsymbol{\omega} + \rho(\boldsymbol{\theta} - \boldsymbol{\vartheta}).
\end{equation}
Finally, by alternatively updating $\boldsymbol{\theta}$, $\boldsymbol{\vartheta}$, and $\boldsymbol{\omega}$, we can solve problem (\ref{pr:Theta_ADMM_AL}) in an iterative manner.

\textit{3) Update RIS reflection amplification factors $\mathbf{a}$:} With obtained $\boldsymbol{\mu}$, $\boldsymbol{\eta}$, $\mathbf{w}_k$ and $\boldsymbol{\theta}$, the optimization problem of designing RIS reflection amplification factors $\mathbf{a}$ can be written as
\begin{equation}\label{pr:A}
\begin{aligned}
\max_\mathbf{a} ~&~~f_2(\mathbf{a})\\
\text{s.t.} ~~& \sum\nolimits_{k=1}^{K}\|\mathbf{\Psi} \mathbf{G} \mathbf{w}_{k}\|^{2}+\|\mathbf{\Psi}\|^{2}_F \sigma_{z}^{2} \leq {P}_{\mathrm{RIS}},\\
& ~a_l \geq 0,~ \forall l.
\end{aligned}
\end{equation}
Similar to the optimization problem of reflection phase-shift coefficients $\boldsymbol{\theta}$, the amplification factors $\mathbf{a}$ is also embedded in the problem (\ref{pr:A}). Thus, we propose to convert it into a more explicit form via a series of matrix manipulations and solve it with an efficient algorithm.

To facilitate the algorithm development, we first recall $\tilde{\boldsymbol{\theta}}_l \triangleq [\theta_{(l-1)Q+1}, \cdots, \theta_{lQ}]^T$ and define $\mathbf{\Phi}_l \triangleq \tilde{\boldsymbol{\theta}}_l\tilde{\boldsymbol{\theta}}_l^T$. Thus, the RIS reflection beamforming $\mathbf{\Psi}$ can be re-denoted as $\mathbf{\Psi} \triangleq \operatorname{blkdiag} \left( \frac{1}{\sqrt{Q}}a_1\mathbf{\Phi}_1, \cdots, \frac{1}{\sqrt{Q}}a_L\mathbf{\Phi}_L \right)$. Utilizing $\mathbf{h}_{\mathrm{r},k,l} \triangleq \mathbf{h}_{\mathrm{r},k}((l-1)Q+1:lQ)$ and $\mathbf{g}_{k,l} \triangleq \mathbf{g}_{k}((l-1)Q+1:lQ)$ to represent vectors consisting of the $[(l-1)Q+1]$-th to the $lQ$-th elements of $\mathbf{h}_{\mathrm{r},k}$ and $\mathbf{g}_{k}$ correspondingly, problem (\ref{pr:A}) can be concisely re-arranged as the following form
\begin{equation}\label{pr:final_A}
\begin{aligned}
\max_{\mathbf{a}} ~~~&\Re\{\mathbf{d}^H\mathbf{a}\} - \mathbf{a}^H\mathbf{R}\mathbf{a}\\
\text{s.t.} ~~~~& \mathbf{a}^H\mathbf{T}\mathbf{a} \leq {P}_{\mathrm{RIS}},\\
& a_l \geq 0,~ \forall l,
\end{aligned}
\end{equation}
where for ease of notation, we define
\begin{subequations}\label{eq:A_trans6}
\begin{align}
    \mathbf{d}^H &\hspace{-0.05cm}\triangleq\hspace{-0.05cm} \sum_{k = 1}^K  2 \sqrt{1+\mu_{k}} \eta_k^{*}\mathbf{b}_{k,k}^H - \sum_{k = 1}^K |\eta_k|^2 \sum_{i = 1}^K\mathbf{c}_{k,i}^H,\\
    \mathbf{b}_{k,i} &\hspace{-0.05cm}\triangleq\hspace{-0.1cm} \frac{1}{\sqrt{Q}}[ \mathbf{g}_{i,1}^H \mathbf{\Phi}_1^H\mathbf{h}_{\mathrm{r},k,1}, \hspace{-0.05cm}\cdots\hspace{-0.05cm}, \mathbf{g}_{i,L}^H \mathbf{\Phi}_L^H\mathbf{h}_{\mathrm{r},k,L}]^T\hspace{-0.1cm}, ~\forall k,\forall i,\\
    \mathbf{c}_{k,i} & \hspace{-0.05cm}\triangleq\hspace{-0.05cm} 2\mathbf{h}_{\mathrm{d},k}^H \mathbf{w}_i\mathbf{b}_{k,i},~ \forall k,\forall i,\\
    \mathbf{R} &\hspace{-0.05cm}\triangleq\hspace{-0.05cm} \sum_{k = 1}^K |\eta_k|^2 \sum_{i = 1}^K \mathbf{b}_{k,i}\mathbf{b}_{k,i}^H + \sum_{k = 1}^K \left|\eta_{k}\right|^{2} \mathbf{S}_k,\\
    \mathbf{S}_k &\hspace{-0.05cm}\triangleq\hspace{-0.05cm} \mathrm{diag}(\mathbf{s}_k), ~\mathbf{s}_k(l) \hspace{-0.05cm}\triangleq\hspace{-0.05cm} \frac{1}{Q} \|\mathbf{h}_{\mathrm{r}, k,l}^{H} \mathbf{\Phi}_l\|^2\sigma_{z}^{2}, ~ \forall k, \\
    \mathbf{T} &\hspace{-0.05cm}\triangleq\hspace{-0.05cm} \mathrm{diag}(\mathbf{t}),~\mathbf{t}(l) \hspace{-0.05cm}\triangleq\hspace{-0.05cm} \frac{1}{Q}\hspace{-0.05cm} \sum_{k = 1}^{K}\hspace{-0.05cm}\| \mathbf{\Phi}_l \mathbf{g}_{k,l} \|^2 \hspace{-0.05cm}+\hspace{-0.05cm}\frac{1}{Q} \|\mathbf{\Phi}_l\|^2_F\sigma_z^2.
\end{align}
\end{subequations}
Note that the problem (\ref{pr:final_A}) is a convex problem whose optimal solution can be obtained easily.

\subsection{Summary}
\label{sec:sum_rate_max}
Through the above analysis, the joint transmit precoding and RIS reflection beamforming design for sum-rate maximization problem is summarized in Algorithm \ref{Algorithm for sum-rate}. With appropriate initializations, the auxiliary variables $\boldsymbol{\mu}$ and $\boldsymbol{\eta}$, the transmit precoding $\mathbf{w}_k$, the RIS reflection phase-shift coefficients $\boldsymbol{\theta}$, and the RIS reflection amplification factors $\mathbf{a}$ are iteratively updated.
Note that the ADMM algorithm is applied to non-convex sub-problem (\ref{pr:Theta}) with respect to $\boldsymbol{\theta}$ with both quadratic constraint and constant-modulus constraint, the theoretical convergence analysis of the ADMM algorithm and the proposed complete Algorithm \ref{Algorithm for sum-rate} cannot be easily obtained. Nevertheless, the simulation results shown in Sec. \ref{simulation} illustrate that the proposed algorithm will converge with an acceptable speed.

\begin{algorithm}[t]
  \caption{Joint Transmit Precoding and RIS Reflection Beamforming Design for Sum-Rate Maximization Problem}
  \label{Algorithm for sum-rate}
  \begin{algorithmic}[1]
    \REQUIRE $\mathbf{h}_{\mathrm{d},k}^H$, $\mathbf{h}_{\mathrm{r},k}^H$, $\mathbf{G}$, $\sigma_k$, $\sigma_\mathrm{z}$, $P_\mathrm{BS}$, $P_\mathrm{RIS}$, $\forall k$.
    \ENSURE  $\mathbf{w}_k^{\star}$, $\boldsymbol{\theta}^{\star}$ and $\mathbf{a}^{\star}$.
    \STATE {Initialize $\mathbf{w}_k$, $\boldsymbol{\theta}$, $\mathbf{a}$, $\boldsymbol{\vartheta}$, and $\boldsymbol{\omega}$.}
    \REPEAT
    \STATE {Update $\boldsymbol{\mu}$ by (\ref{eq:optimal_mu});}
    \STATE {Update $\boldsymbol{\eta}$ by (\ref{eq:optimal_eta});}
    \STATE {Update $\mathbf{w}_k$ by solving (\ref{pr:final_w});}
    \REPEAT
    \STATE {Update $\boldsymbol{\theta}$ by solving  (\ref{pr:theta_final});}
    \STATE {Update $\boldsymbol{\vartheta}$ by (\ref{eq:vartheta});}
    \STATE {Update $\boldsymbol{\omega}$ by (\ref{eq:omega});}
    \UNTIL {convergence}.
    \STATE {Update $\mathbf{a}$ by solving (\ref{pr:final_A});}
    \UNTIL {convergence}.
    \STATE {Return $\mathbf{w}_k^{\star} = \mathbf{w}_k$, $\boldsymbol{\theta}^{\star} = \boldsymbol{\theta}$ and $\mathbf{a}^{\star} = \mathbf{a}$}.
  \end{algorithmic}
\end{algorithm}

\textit{1) Initialization:} For the above BCD algorithm, especially the ADMM-MM based algorithm for designing $\boldsymbol{\theta}$, a good initial point can accelerate the convergence and promote the performance. Therefore, in the following, we investigate to appropriately initialize the RIS reflection phase-shift coefficients $\boldsymbol{\theta}$, the RIS amplification factors $\mathbf{a}$, and the transmit precoding $\mathbf{w}_k, \forall k$.

In general, RIS is deployed to improve the quality of channels and create more favorable radio environments. Thus, we utilize channel power gain $\sum\nolimits_{k = 1}^K\|\mathbf{h}_{\mathrm{d},k}^H + \mathbf{h}_{\mathrm{r},k}^H\mathbf{\Psi}\mathbf{G} \|^2$ as the metric to initialize the RIS reflection phase-shift coefficients. To simplify the initialization problem, we assume $L = M$ and RIS amplification factors $\mathbf{a} = \mathbf{1}_{L}$, i.e., $\mathbf{\Xi} = \mathbf{I}_{M \times M}$. Thus, the RIS reflection beamforming $\mathbf{\Psi}$ can be re-written by $\mathbf{\Psi} = \mathbf{\Theta}\mathbf{\Xi}\mathbf{\Theta} = \mathbf{\Theta}^2$. Obviously, the channel power gain is a quadratic function on $\mathbf{\Theta}$, which is challenging to optimize directly. Thus, we attempt to seek for a simpler approach to solve it. Considering the relationship between $\mathbf{\Psi}$ and $\mathbf{\Theta}$, we first optimize the more tractable problem on variable $\mathbf{\Psi}$ and then obtain the initial $\mathbf{\Theta}$ by a series of angular operations. In details, we define $\boldsymbol{\psi} = [\psi_1, \cdots, \psi_M]^T$ as a constant-modulus vector consisting of the diagonal elements of the diagonal matrix $\mathbf{\Psi}$, i.e., $\mathbf{\Psi} \triangleq \mathrm{diag}{(\boldsymbol{\psi})}$. The optimization problem for $\mathbf{\Psi}$ is given by
\begin{equation}\label{pr:initial_psi}
\begin{aligned}
\max_{\boldsymbol{\psi}} ~&~\sum\nolimits_{k = 1}^K\|\mathbf{h}_{\mathrm{d},k}^H + \mathbf{h}_{\mathrm{r},k}^H\mathbf{\Psi}\mathbf{G} \|^2\\
\text{s.t.} ~~& ~~|\psi_m| = 1,~ \forall m.
\end{aligned}
\end{equation}
Then, by defining
\begin{subequations}\label{eq:ini_trans}
\begin{align}
\breve{\boldsymbol{\psi}} &\triangleq \boldsymbol{\psi}^{*},\\
\breve{\mathbf{R}}_k &\triangleq \mathrm{diag}(\mathbf{h}_{\mathrm{r},k}^*)\mathbf{G},\\
\mathbf{M} &\triangleq \sum\nolimits_{k = 1}^K \breve{\mathbf{R}}_k\breve{\mathbf{R}}_k^H,\\
\mathbf{m} &\triangleq 2\sum\nolimits_{k = 1}^K\breve{\mathbf{R}}_k\mathbf{h}_{\mathrm{d},k},
\end{align}
\end{subequations}
problem (\ref{pr:initial_psi}) can be concisely re-formulated as
\begin{equation}\label{pr:initial_final}
\begin{aligned}
\max_{\breve{\boldsymbol{\psi}}} ~&~f_3(\breve{\boldsymbol{\psi}}) \triangleq \breve{\boldsymbol{\psi}}^H\mathbf{M}\breve{\boldsymbol{\psi}} + \Re\{\breve{\boldsymbol{\psi}}^H\mathbf{m}\} \\
\text{s.t.} ~~& ~|\breve{\psi}_m| = 1,~ \forall m,
\end{aligned}
\end{equation}
which can be effectively solved by utilizing popular Riemannian conjugate gradient (RCG) algorithm \cite{LiuR}, \cite{LiuR2}. The details of RCG algorithm are omitted due to space limitations. After obtaining $\breve{\boldsymbol{\psi}}$, we can construct $\boldsymbol{\psi}$ by $\boldsymbol{\psi} = \breve{\boldsymbol{\psi}}^{*}$. Then, recalling $\mathbf{\Psi} = \mathbf{\Theta}^2$, i.e., $\boldsymbol{\psi} = \boldsymbol{\theta}^2$, we can easily obtain the initial $\boldsymbol{\theta}$ by $\angle{\theta_m} = \angle{\psi_m}/2, ~\forall m$. Moreover, we can simply initialize the RIS reflection factors $\mathbf{a} = \mathbf{1}_{L}$ and transmit precoding $\mathbf{w}_k = \frac{\sqrt{P_\mathrm{BS}}(\sum\nolimits_{k = 1}^K \mathbf{h}_k\mathbf{h}_k^H + \tilde{\sigma}_k^2\mathbf{I})^{-1}\mathbf{h}_k}{\sqrt{K}\|\sum\nolimits_{k = 1}^K \mathbf{h}_k\mathbf{h}_k^H + \tilde{\sigma}_k^2\mathbf{I})^{-1}\mathbf{h}_k\|},~\forall k$, according to the popular minimum mean squared error (MMSE) criterion, where $\tilde{\sigma}_k^2 \triangleq \|\mathbf{h}_{\mathrm{r}, k}^{H} \mathbf{\Psi}\|^{2} \sigma_{z}^{2}+\sigma_k^{2}$.

\textit{2) Computational Complexity Analysis:}
We assume that the popular interior point method is adopted to solve convex problems, whose complexity is relevant to the dimension of the variable and the number of linear matrix inequality (LMI) constraints and second-order cone (SOC) constraints \cite{WangKY}. In the initialization stage, it requires at most $\mathcal{O}({M^{1.5}})$ operations to obtain the phase-shift vector $\boldsymbol{\theta}$ \cite{Absil}. Besides, the complexities for obtaining initial $\mathbf{w}_k$ and $\mathbf{a}$ are of order $\mathcal{O}({KN^3})$ and $\mathcal{O}({L})$. In each outer iteration, obtaining the optimal solutions of $\boldsymbol{\mu}$ and $\boldsymbol{\eta}$ requires approximately $\mathcal{O}({K(NK+M^2)})$ and $\mathcal{O}({K[N(K+1)+M^2]})$, respectively; updating $\mathbf{w}_k$ requires about $\mathcal{O}(\sqrt{NK+2}(1+NK)N^3K^3)$ operations; the calculation of $\mathbf{a}$ has a computational complexities of $\mathcal{O}(\sqrt{L+1}(1+L)L^3)$. The computational complexity of proposed ADMM-MM-based algorithm for designing $\boldsymbol{\theta}$ lies in the updates of three variables $\boldsymbol{\theta}$, $\boldsymbol{\vartheta}$ and $\boldsymbol{\omega}$. Solving sub-problem with respect to $\boldsymbol{\theta}$ has the complexities of order $\mathcal{O}(\sqrt{M+1}(1+M)M^3)$.  The complexity of updating the closed-form $\boldsymbol{\vartheta}$ is of order $\mathcal{O}(M)$. The complexity to update the dual variable $\boldsymbol{\omega}$ is of order $\mathcal{O}(M)$. To sum up, the total computational complexity of Algorithm 1 is approximated at the order of $\mathcal{O}(I_\mathrm{o}[\sqrt{NK+2}(1+NK)N^3K^3 + I_{\mathrm{i},s}(\sqrt{M+1}(1+M)M^3) + \sqrt{L+1}(1+L)L^3])$, in which $I_\mathrm{o}$ and $I_{\mathrm{i},s}$ are the number of iterations required for convergence of the outer loop and the inner ADMM-MM loop in the $s$-th outer loop, respectively.

\section{Joint Design for Power Minimization Problem}
\subsection{Problem Formulation}
In this section, our goal is to design transmit precoding $\mathbf{w}_k$, RIS reflection phase-shift coefficients $\boldsymbol{\theta}$ and RIS reflection amplification factors $\mathbf{a}$ to minimize the total power consumption at the BS and the sub-connected active RIS subject to users' SINR requirements. This power minimization problem can be formulated as
\begin{subequations}\label{pr:original_problem_p}
\begin{align}
\hspace{-0.5cm}\min_{\mathbf{w}_k, \boldsymbol{\theta}, \mathbf{a}}  ~~&~ \mathcal{P}_\mathrm{b} + \mathcal{P}_\mathrm{r}\\
\text{s.t.} ~~~~&~ \gamma_{k} \geq \Gamma_{k},~  \forall k,\\
&~ |\theta_m| = 1,~ \forall m, \\
&~ a_l \geq 0,~ \forall l,
\end{align}
\end{subequations}
where we recall that $\mathcal{P}_\mathrm{b} = \nu_1^{-1}\sum\nolimits_{k=1}^{K}\|\mathbf{w}_{k}\|^{2} + W_\mathrm{BS}$ is the power consumption at the BS and  $\mathcal{P}_\mathrm{r} =  \nu_2^{-1}\Big(\sum\nolimits_{k=1}^{K}\|\mathbf{\Psi} \mathbf{G} \mathbf{w}_{k}\|^{2}+\|\mathbf{\Psi}\|^{2}_F \sigma_{z}^{2}\Big) + M W_\mathrm{PS} + L W_\mathrm{PA}$ is the power consumption at the sub-connected active RIS.
Moreover, (\ref{pr:original_problem_p}b) represents users' quality-of-service (QoS) constraints, in which $\Gamma_{k} > 0$ is the minimum SINR requirement of the user-$k$.
Note that problem (\ref{pr:original_problem_p}) is still challenging to solve since the transmit precoding $\mathbf{w}_k$, RIS reflection coefficients $\boldsymbol{\theta}$ and $\mathbf{a}$ are tightly coupled in the QoS constraints. To address these issues, in the following subsection, we propose to decouple the original problem and iteratively optimize $\mathbf{w}_k$, $\boldsymbol{\theta}$ and $\mathbf{a}$.
\subsection{Block update}
\textit{1) Update precoding $\mathbf{w}_k$:} When RIS reflection coefficients $\boldsymbol{\theta}$ and $\mathbf{a}$ are fixed, the sub-problem of optimizing transmit precoding $\mathbf{w}_k$ is reduced to
\begin{equation}\label{pr:pm_w}
\begin{aligned}
\min_{\mathbf{w}_k} ~~&~~~\nu_1^{-1}\sum\nolimits_{k=1}^{K}\|\mathbf{w}_{k}\|^{2} + \nu_2^{-1}\sum\nolimits_{k=1}^{K}\|\mathbf{\Psi} \mathbf{G} \mathbf{w}_{k}\|^{2}\\
\text{s.t.} ~~~&~~~ \gamma_{k} \geq \Gamma_{k},~  \forall k,
\end{aligned}
\end{equation}
which is a conventional power minimization problem and can be efficiently solved by using second-order cone program (SOCP) algorithm.

\textit{2) Update RIS reflection phase-shift coefficients $\boldsymbol{\theta}$:} When the other variables are determined, the optimization problem of solving for the RIS reflection phase-shift coefficients $\boldsymbol{\theta}$ can be simplified as
\begin{equation}\label{pr:pm_theta_pre}
  \min_{\boldsymbol{\theta}} ~~~ \sum\nolimits_{k=1}^{K}\|\mathbf{\Psi} \mathbf{G} \mathbf{w}_{k}\|^{2},
\end{equation}
where we should emphasize that the equation $\|\mathbf{\Psi}\|^{2}_F\sigma_{z}^{2} = \|\mathbf{\Xi}\|^{2}_F\sigma_{z}^{2}$ is independent of $\boldsymbol{\theta}$ and so we ignore it in (\ref{pr:pm_theta_pre}). Furthermore, problem (\ref{pr:pm_theta_pre}) is equivalent to
\begin{subequations}\label{pr:pm_theta}
\begin{align}
\min_{\boldsymbol{\theta}} ~~&~ \boldsymbol{\theta}^H\mathbf{Q}_2\boldsymbol{\theta} \\
\text{s.t.} ~~~&~ \gamma_{k} \geq \Gamma_{k},~  \forall k,\\
&~ |\theta_m| = 1,~ \forall m,
\end{align}
\end{subequations}
where we recall $\mathbf{Q}_2 \triangleq \sum\nolimits_{k = 1}^K \operatorname{diag}(\mathbf{g}_{k}^*)\mathbf{\Xi}^H \mathbf{\Xi}\operatorname{diag}(\mathbf{g}_{k})$ as defined in (\ref{eq:theta_trans4}h). Due to $K$ non-convex constraints (\ref{pr:pm_theta}b) and unit modulus constraint (\ref{pr:pm_theta}c), it is very challenging to directly seek for an optimal solution to problem (\ref{pr:pm_theta}). Therefore, in the follows, we also utilize the similar ADMM-MM-based algorithmic framework to tackle these issues. First, in order to facilitate the development of ADMM algorithm, we introduce an auxiliary variable $\boldsymbol{\vartheta}$ to convert (\ref{pr:pm_theta}) into
\begin{subequations}\label{pr:pm_theta_ADMM}
\begin{align}
\min_{\boldsymbol{\theta}, \boldsymbol{\vartheta}} ~~&~ \boldsymbol{\theta}^H\mathbf{Q}_2\boldsymbol{\theta} \\
\text{s.t.} ~~~& \mathbf{v}^H \widehat{\mathbf{F}}_{k}\mathbf{v} + \Re\{\boldsymbol{\theta}^H\widehat{\mathbf{P}}_{k}\boldsymbol{\theta}^{*}\} + \boldsymbol{\theta}^H \widehat{\mathbf{Q}}_{k}\boldsymbol{\theta} + \varsigma_k \leq 0 ,~  \forall k,\\
& |\theta_m| \leq 1,~ \forall m,\\
& \boldsymbol{\vartheta} = \boldsymbol{\theta},\\
& |\varsigma_m| = 1,~ \forall m,
\end{align}
\end{subequations}
where for simplicity we define
\begin{subequations}\label{eq:pm_theta_trans}
\begin{align}
  \widehat{\mathbf{F}}_{k} &\triangleq \Gamma_{k} \sum\nolimits_{ i \neq k}^{K} \mathbf{F}_{k,i} - \mathbf{F}_{k,k},~ \mathbf{F}_{k,i} \triangleq \mathbf{f}_{k,i}\mathbf{f}_{k,i}^H,\\
  \widehat{\mathbf{P}}_{k} &\triangleq 2\Gamma_{k} \sum\nolimits_{ i \neq k}^{K}(\mathbf{h}_{\mathrm{d},k}^H\mathbf{w}_i) \tilde{\mathbf{P}}_{k,i}^T - 2 (\mathbf{h}_{\mathrm{d},k}^H\mathbf{w}_k)\tilde{\mathbf{P}}_{k,k}^T,\\
  \widehat{\mathbf{Q}}_{k} &\triangleq \Gamma_{k}(\operatorname{diag}(\mathbf{h}_{\mathrm{r}, k}^*)\mathbf{\Xi}\mathbf{\Xi}^H \operatorname{diag}(\mathbf{h}_{\mathrm{r}, k}))^{*}\sigma_z^2,\\
  \varsigma_{k} &\triangleq \Gamma_{k} \sum\nolimits_{ i \neq k}^{K} |\mathbf{h}_{\mathrm{d},k}^H\mathbf{w}_i|^2 + \Gamma_{k}\sigma_k^2 - |\mathbf{h}_{\mathrm{d},k}^H\mathbf{w}_k|^2.
\end{align}
\end{subequations}
For problem (\ref{pr:pm_theta_ADMM}), its augmented Lagrangian problem can be formulated as
\begin{subequations}\label{pr:pm_theta_ADMM_AL}
\begin{align}
\min_{\boldsymbol{\theta},\boldsymbol{\vartheta}, \boldsymbol{\omega}} ~~& \boldsymbol{\theta}^H\mathbf{Q}_2\boldsymbol{\theta} +  \frac{\rho}{2}\|\boldsymbol{\theta} - \boldsymbol{\vartheta} + \frac{\boldsymbol{\omega}}{\rho}\|^2\\
\text{s.t.} ~~~~& \mathbf{v}^H \widehat{\mathbf{F}}_{k}\mathbf{v} + \Re\{\boldsymbol{\theta}^H\widehat{\mathbf{P}}_{k}\boldsymbol{\theta}^{*}\} + \boldsymbol{\theta}^H \widehat{\mathbf{Q}}_{k}\boldsymbol{\theta} + \varsigma_k \leq 0 ,~  \forall k,\\
& |\theta_m| \leq 1,~ \forall m,\\
& |\varsigma_m| = 1,~ \forall m,
\end{align}
\end{subequations} where we introduce penalty parameter $\rho$ and dual variable $\boldsymbol{\omega}$. Obviously, it can be effectively tackled by alternately optimizing $\boldsymbol{\theta}$, $\boldsymbol{\vartheta}$, and $\boldsymbol{\omega}$. The details of the algorithm will be presented in the following.

\textbf{Update $\boldsymbol{\theta}$:} To deal with the non-convex terms $\mathbf{v}^H \widehat{\mathbf{F}}_{k}\mathbf{v}$ and $\Re\{\boldsymbol{\theta}^H\widehat{\mathbf{P}}_{k}\boldsymbol{\theta}^{*}\}$ in constraint (\ref{pr:pm_theta_ADMM_AL}b), we employ the similar MM procedure introduced in the previous section to seek a series of tractable surrogate functions for them. In particular, with the solution $\boldsymbol{\theta}_t$ obtained in the previous iteration and the definitions $\mathbf{F}_{k,1} \triangleq \Gamma_{k} \sum\nolimits_{ i \neq k}^{K} \mathbf{F}_{k,i}$ and $\mathbf{F}_{k,2} \triangleq -\mathbf{F}_{k,k}$, where $\widehat{\mathbf{F}}_{k} = \mathbf{F}_{k,1} + \mathbf{F}_{k,2}$, approximate upper-bounds for the first term $\mathbf{v}^H\mathbf{F}_{k,1}\mathbf{v}$ and the second term $\mathbf{v}^H\mathbf{F}_{k,2}\mathbf{v}$ of $\mathbf{v}^H \widehat{\mathbf{F}}_{k}\mathbf{v}$ can be constructed via the second-order Taylor expansion as:
\begin{subequations}\label{eq:pm_theta_MM1_0}
\begin{align}
    \mathbf{v}^H\mathbf{F}_{k,1}\mathbf{v}  &\leq  \lambda_{k,1} \mathbf{v}^H\mathbf{v} + 2\Re\{\mathbf{v}^H(\mathbf{F}_{k,1} - \lambda_{k,1}\mathbf{I}_{M^2})\mathbf{v}_{t}\} \notag\\
    &\quad + \mathbf{v}_t^H(\lambda_{k,1}\mathbf{I}_{M^2} - {\mathbf{F}}_{k,1})\mathbf{v}_{t},\\
    \mathbf{v}^H\mathbf{F}_{k,2}\mathbf{v}  &\leq  \lambda_{k,2} \mathbf{v}^H\mathbf{v} + 2\Re\{\mathbf{v}^H(\mathbf{F}_{k,2} - \lambda_{k,2}\mathbf{I}_{M^2})\mathbf{v}_{t}\} \notag\\
    &\quad+ \mathbf{v}_t^H(\lambda_{k,2}\mathbf{I}_{M^2} - {\mathbf{F}}_{k,2})\mathbf{v}_{t},
\end{align}
\end{subequations}
where we choose
\begin{subequations}\label{eq:pm_theta_MM_lk}
\begin{align}
  \lambda_{k,1} &= \mathrm{Tr}\Big\{\Gamma_{k} \sum\nolimits_{ i \neq k}^{K} \mathbf{F}_{k,i}\Big\} = \Gamma_{k} \sum\nolimits_{ i \neq k}^{K} \|\mathbf{f}_{k,i}\|^2, \\
  \lambda_{k,2} &= \mathrm{Tr}\Big\{ \mathbf{F}_{k,k}\Big\} = \|\mathbf{f}_{k,k}\|^2,
\end{align}
\end{subequations}
as efficient solutions to avoid the ultra-high computational complexity of matrix calculation and eigenvalue decomposition. Consequently, the upper-bound function of $\mathbf{v}^H \widehat{\mathbf{F}}_{k}\mathbf{v}$ can be derived as
\begin{equation}\label{eq:pm_theta_MM1}
\begin{aligned}
    \mathbf{v}^H\widehat{\mathbf{F}}_{k}\mathbf{v} &= \mathbf{v}^H\mathbf{F}_{k,1}\mathbf{v} + \mathbf{v}^H\mathbf{F}_{k,2}\mathbf{v}\\
    &\leq  \lambda_k \mathbf{v}^H\mathbf{v} + 2\Re\{\mathbf{v}^H(\widehat{\mathbf{F}}_{k} - \lambda_k\mathbf{I}_{M^2})\mathbf{v}_{t}\}\\
    &\quad + \mathbf{v}_t^H(\lambda_k\mathbf{I}_{M^2} - \widehat{\mathbf{F}}_{k})\mathbf{v}_{t},\\
\end{aligned}
\end{equation}
with $\lambda_k \triangleq \lambda_{k,1} + \lambda_{k,2}$.

Meanwhile, recalling $\mathbf{v}^H\mathbf{v} \leq M^2$, we can re-construct the upper-bound surrogate function of $\mathbf{v}^H\widehat{\mathbf{F}}_{k}\mathbf{v}$ as
\begin{equation}\label{eq:pm_theta_MM1_1}
\begin{aligned}
    \mathbf{v}^H\widehat{\mathbf{F}}_{k}\mathbf{v} &\leq \Re\{\mathbf{v}^H\widehat{\mathbf{f}}_{{k,t}}\} +c_{k,t}=   \Re\{\boldsymbol{\theta}^H\widehat{\mathbf{F}}_{{k,t}} \boldsymbol{\theta}^{*}\} +c_{k,t},
\end{aligned}
\end{equation}
in which
\begin{subequations}\label{eq:pm_theta_MM_ck}
\begin{align}
 \widehat{\mathbf{f}}_{{k,t}} &\triangleq 2(\widehat{\mathbf{F}}_{k} - \lambda_k\mathbf{I}_{M^2})\mathbf{v}_{t} = \mathrm{vec}\{\widehat{\mathbf{F}}_{{k,t}}\}, \\
 \widehat{\mathbf{F}}_{{k,t}} &\triangleq 2\Gamma_{k}\hspace{-0.1cm}\sum_{i \neq k}^K  \hspace{-0.05cm}\boldsymbol{\theta}_{t}^T\hspace{-0.025cm}\tilde{\mathbf{P}}_{k,i}^H\boldsymbol{\theta}_{t}\tilde{\mathbf{P}}_{k,i}\hspace{-0.05cm}-\hspace{-0.05cm}2\boldsymbol{\theta}_{t}^T\hspace{-0.025cm}\tilde{\mathbf{P}}_{k,k}^H\boldsymbol{\theta}_{t}\tilde{\mathbf{P}}_{k,k}\hspace{-0.05cm}- \hspace{-0.05cm}2\lambda_k\boldsymbol{\theta}_{t}\boldsymbol{\theta}_{t}^T,\\
  c_{k,t} &\triangleq 2\lambda_kM^2 - \Gamma_{k} \sum\nolimits_{ i \neq k}^{K} |\mathbf{v}_t^H \mathbf{f}_{k,i}|^2 + |\mathbf{v}_t^H \mathbf{f}_{k,k}|^2.
\end{align}
\end{subequations}

Therefore, plugging the result in (\ref{eq:pm_theta_MM1_1}) into (\ref{pr:pm_theta_ADMM_AL}b), we can re-formulate the users' SINR constraints in each iteration as
\begin{equation}\label{eq:pm_MM}
   \Re\{\boldsymbol{\theta}^H\widehat{\mathbf{P}}_{{k,t}} \boldsymbol{\theta}^{*}\} + \boldsymbol{\theta}^H \widehat{\mathbf{Q}}_{k}\boldsymbol{\theta} + \varsigma_k +c_{k,t} \leq 0,~ \forall k,
\end{equation}
where we define $\widehat{\mathbf{P}}_{{k,t}} \triangleq \widehat{\mathbf{F}}_{{k,t}} + \widehat{\mathbf{P}}_{k}$. Notice that $\Re\{\boldsymbol{\theta}^H\widehat{\mathbf{P}}_{{k,t}} \boldsymbol{\theta}^{*}\}$ is still a non-convex real-valued function and challenging to deal with. By defining
\begin{equation}\label{eq:real_var2}
\bar{\mathbf{P}}_{{k,t}} \triangleq \begin{bmatrix}
\Re\{\widehat{\mathbf{P}}_{{k,t}}\} & \Im\{\widehat{\mathbf{P}}_{{k,t}}\} \\
\Im\{\widehat{\mathbf{P}}_{{k,t}}\} & -\Re\{\widehat{\mathbf{P}}_{{k,t}}\}
\end{bmatrix},
\end{equation}
and recalling $\bar{\boldsymbol{\theta}} \triangleq \left [\Re\{\boldsymbol{\theta}^{T}\} \ \ \Im\{\boldsymbol{\theta}^{T}\}\right ]^{T}$, we can convert the complex-valued variables $\Re\{\boldsymbol{\theta}^H \widehat{\mathbf{P}}_{{k,t}}  \boldsymbol{\theta}^{*}\} $ into real-valued ones $\bar{\boldsymbol{\theta}}^T \bar{\mathbf{P}}_{{k,t}}\bar{\boldsymbol{\theta}}$, and then apply the idea of MM again to seek a series of tractable surrogate functions for it. In particular, the upper-bound function can be constructed by utilizing the second-order Taylor expansion as
\begin{subequations}\label{eq:pm_theta_MM2}
\begin{align}
  \Re\{\boldsymbol{\theta}^H \widehat{\mathbf{P}}_{k,t}  \boldsymbol{\theta}^{*}\} & = \bar{\boldsymbol{\theta}}^T \bar{\mathbf{P}}_{{k,t}}\bar{\boldsymbol{\theta}},\\
  & \leq \bar{\boldsymbol{\theta}}_t^T\bar{\mathbf{P}}_{{k,t}} \bar{\boldsymbol{\theta}}_t + \bar{\boldsymbol{\theta}}_t^T(\bar{\mathbf{P}}_{{k,t}} + \bar{\mathbf{P}}_{{k,t}}^T)(\bar{\boldsymbol{\theta}} - \bar{\boldsymbol{\theta}}_t) \notag\\
  & \quad + \frac{\lambda_{\mathrm{q},k,t}}{2}(\bar{\boldsymbol{\theta}} - \bar{\boldsymbol{\theta}}_t)^T(\bar{\boldsymbol{\theta}} - \bar{\boldsymbol{\theta}}_t),\\
  & = \frac{\lambda_{\mathrm{q},k,t}}{2} \bar{\boldsymbol{\theta}}^T\bar{\boldsymbol{\theta}} + \Re\{\bar{\boldsymbol{\theta}}^T \bar{\mathbf{q}}_{k,t}\} + c_{\mathrm{q},k,t},\\
  & = \frac{\lambda_{\mathrm{q},k,t}}{2}\boldsymbol{\theta}^H\boldsymbol{\theta} + \Re\{\boldsymbol{\theta}^H \mathbf{U}\bar{\mathbf{q}}_{k,t}\}+ c_{\mathrm{q},k,t},
\end{align}
\end{subequations}
where $\lambda_{\mathrm{q},k,t}$ is the maximum eigenvalue of the Hessian matrix $(\bar{\mathbf{P}}_{{k,t}} + \bar{\mathbf{P}}_{{k,t}}^T)$, $\bar{\mathbf{q}}_{k,t} \triangleq (\bar{\mathbf{P}}_{{k,t}} + \bar{\mathbf{P}}_{{k,t}}^T - \lambda_{\mathrm{q},k,t}\mathbf{I}_{2M})\bar{\boldsymbol{\theta}}_t$, and $c_{\mathrm{q},k,t} = \bar{\boldsymbol{\theta}}_t^T \bar{\mathbf{P}}_{{k,t}} \bar{\boldsymbol{\theta}}_t - \bar{\boldsymbol{\theta}}_t^T(\bar{\mathbf{P}}_{{k,t}} + \bar{\mathbf{P}}_{{k,t}}^T)\bar{\boldsymbol{\theta}}_t + \frac{\lambda_{\mathrm{q},k,t}}{2}\bar{\boldsymbol{\theta}}_t^T\bar{\boldsymbol{\theta}}_t$ is a constant independent of $\boldsymbol{\theta}$.
In summary, an appropriate surrogate function for the function on the left side of (\ref{pr:pm_theta_ADMM_AL}b) can be expressed as
\begin{equation}\label{eq:pm_theta_final_MM+MM}
\begin{aligned}
  \widehat{g}_k(\boldsymbol{\theta}|\boldsymbol{\theta}_t)& = \boldsymbol{\theta}^H \widehat{\mathbf{Q}}_{k}\boldsymbol{\theta} + \frac{\lambda_{\mathrm{q},k,t}}{2}\boldsymbol{\theta}^H\boldsymbol{\theta} + \Re\{\boldsymbol{\theta}^H \mathbf{U}\bar{\mathbf{q}}_{k,t}\}+ c_{\mathrm{e},k,t},\\
  & = \boldsymbol{\theta}^H \mathbf{\Lambda}_{k,t}\boldsymbol{\theta} + \Re\{\boldsymbol{\theta}^H\boldsymbol{\beta}_{k,t}\} + c_{\mathrm{e},k,t},
\end{aligned}
\end{equation} where for brevity we define $\mathbf{\Lambda}_{k,t} \triangleq \widehat{\mathbf{Q}}_k +  \frac{\lambda_{\mathrm{q},k,t}}{2} \mathbf{I}_{M}$, $\boldsymbol{\beta}_{k,t} \triangleq \mathbf{U}\bar{\mathbf{q}}_{k,t}$ and $c_{\mathrm{e},k,t} \triangleq c_{\mathrm{q},k,t} + c_{k,t} + \varsigma_k$. Thus, the optimization problem for updating $\boldsymbol{\theta}$ can be formulated as
\begin{equation}\label{pr:pm_theta_ADMM_AL_final}
\begin{aligned}
\min_{\boldsymbol{\theta}} ~~& \boldsymbol{\theta}^H \mathbf{\Omega}\boldsymbol{\theta} + \Re\{\boldsymbol{\theta}^H \boldsymbol{\varrho}\}\\
\text{s.t.} ~~~&  \widehat{g}_k(\boldsymbol{\theta}|\boldsymbol{\theta}_t) \leq 0 ,~  \forall k,\\
& |\theta_m| \leq 1,~ \forall m,
\end{aligned}
\end{equation} in which $\mathbf{\Omega} \triangleq \mathbf{Q}_2 + \frac{\rho}{2} \mathbf{I}_{M}$ and $\boldsymbol{\varrho} \triangleq - \rho \boldsymbol{\vartheta} + \boldsymbol{\omega}$.

\textbf{Update $\boldsymbol{\vartheta}$ and $\boldsymbol{\omega}$:}
Similar to the sum-rate maximization problem, $\boldsymbol{\vartheta}$ and $\boldsymbol{\omega}$ can be updated by (\ref{eq:vartheta}) and (\ref{eq:omega}), respectively.

Now, the ADMM-MM-based algorithm for solving phase-shift $\boldsymbol{\theta}$ optimization problem (\ref{pr:pm_theta}) is straightforward. By sequentially updating $\boldsymbol{\theta}$, $\boldsymbol{\vartheta}$, and $\boldsymbol{\omega}$ until convergence is achieved, we can obtain the optimal RIS reflection phase-shift coefficients $\boldsymbol{\theta}$.

\textit{3) Update RIS reflection amplification factors $\mathbf{a}$:} Fixed transmit precoding $\mathbf{w}_k$ and RIS reflection phase-shift coefficients $\boldsymbol{\theta}$, RIS reflection amplification factors $\mathbf{a}$ can be updated by solving the following problem
\begin{equation}\label{pr:pm_a}
\begin{aligned}
\min_{\mathbf{a}} ~~&~ \sum\nolimits_{k=1}^{K}\|\mathbf{\Psi} \mathbf{G} \mathbf{w}_{k}\|^{2}+\|\mathbf{\Psi}\|^{2}_F \sigma_{z}^{2} \\
\text{s.t.} ~~~&~~ \gamma_{k} \geq \Gamma_{k},~  \forall k,\\
&~~ a_l \geq 0,~ \forall l.
\end{aligned}
\end{equation}
Recalling
\begin{subequations}\label{eq:trans_pm_a}
\begin{align}
    \mathbf{b}_{k,i} &\hspace{-0.05cm}\triangleq\hspace{-0.1cm} \frac{1}{\sqrt{Q}}[ \mathbf{g}_{i,1}^H \mathbf{\Phi}_1^H\mathbf{h}_{\mathrm{r},k,1}, \hspace{-0.05cm}\cdots\hspace{-0.05cm}, \mathbf{g}_{i,L}^H \mathbf{\Phi}_L^H\mathbf{h}_{\mathrm{r},k,L}]^T\hspace{-0.1cm}, ~\forall k,\forall i,\\
    \mathbf{S}_k &\hspace{-0.05cm}\triangleq\hspace{-0.05cm} \mathrm{diag}(\mathbf{s}_k), ~\mathbf{s}_k(l) \hspace{-0.05cm}\triangleq\hspace{-0.05cm} \frac{1}{Q} \|\mathbf{h}_{\mathrm{r}, k,l}^{H} \mathbf{\Phi}_l\|^2\sigma_{z}^{2}, ~ \forall k, \\
    \mathbf{T} &\hspace{-0.05cm}\triangleq\hspace{-0.05cm} \mathrm{diag}(\mathbf{t}),~\mathbf{t}(l) \hspace{-0.05cm}\triangleq\hspace{-0.05cm} \frac{1}{Q}\hspace{-0.05cm} \sum_{k = 1}^{K}\hspace{-0.05cm}\| \mathbf{\Phi}_l \mathbf{g}_{k,l} \|^2 \hspace{-0.05cm}+\hspace{-0.05cm}\frac{1}{Q} \|\mathbf{\Phi}_l\|^2_F\sigma_z^2,
\end{align}
\end{subequations}
as previously defined in (\ref{eq:A_trans6}b), (\ref{eq:A_trans6}e), and (\ref{eq:A_trans6}f), we can re-arrange problem (\ref{pr:pm_a}) as
\begin{equation}\label{pr:pm_a_2}
\begin{aligned}
\min_{\mathbf{a}} ~~&~~~~ \mathbf{a}^H\mathbf{T}\mathbf{a} \\
\text{s.t.} ~~~&~ \frac{|\mathbf{h}_{\mathrm{d},k}^H\mathbf{w}_k + \mathbf{b}_{k,k}^H\mathbf{a}|^2}{\sum\nolimits_{i \neq k}^{K}|\mathbf{h}_{\mathrm{d},k}^H\mathbf{w}_i + \mathbf{b}_{k,i}^H\mathbf{a}|^2 +\mathbf{a}^H\mathbf{S}_k\mathbf{a} + \sigma_k^2} \geq \Gamma_k,~  \forall k,\\
&~~ a_l \geq 0,~ \forall l,
\end{aligned}
\end{equation}
which is a standard SOCP optimization problem and can be effectively solved by CVX.

\subsection{Summary}

\begin{algorithm}[t]
  \caption{Joint Transmit Precoding and RIS Reflection Beamforming Design for Power Minimization Problem}
  \label{Algorithm for power}
  \begin{algorithmic}[1]
    \REQUIRE $\mathbf{h}_{\mathrm{d},k}^H$, $\mathbf{h}_{\mathrm{r},k}^H$, $\mathbf{G}$, $\sigma_k$, $\sigma_\mathrm{z}$, $\Gamma_k$, $\forall k$.
    \ENSURE  $\mathbf{w}_k^{\star}$, $\boldsymbol{\theta}^{\star}$ and $\mathbf{a}^{\star}$.
    \STATE {Initialize $\mathbf{w}_k$, $\boldsymbol{\theta}$, $\mathbf{a}$, $\boldsymbol{\vartheta}$, and $\boldsymbol{\omega}$.}
    \REPEAT
    \STATE {Update $\mathbf{w}_k$ by solving (\ref{pr:pm_w});}
    \REPEAT
    \STATE {Update $\boldsymbol{\theta}$ by solving  (\ref{pr:pm_theta_ADMM_AL_final});}
    \STATE {Update $\boldsymbol{\vartheta}$ by (\ref{eq:vartheta});}
    \STATE {Update $\boldsymbol{\omega}$ by (\ref{eq:omega});}
    \UNTIL {convergence}.
    \STATE {Update $\mathbf{a}$ by solving (\ref{pr:pm_a_2});}
    \UNTIL {convergence}.
    \STATE {Return $\mathbf{w}_k^{\star} = \mathbf{w}_k$, $\boldsymbol{\theta}^{\star} = \boldsymbol{\theta}$ and $\mathbf{a}^{\star} = \mathbf{a}$}.
  \end{algorithmic}
\end{algorithm}

Based on the above derivations, the joint transmit precoding and RIS reflection beamforming design for power minimization problem is straightforward and summarized in Algorithm \ref{Algorithm for power}. The appropriate initializations $\mathbf{w}_k, \forall k$, $\boldsymbol{\theta}$, and $\mathbf{a}$ can be obtained by the similar methods presented in Sec. \ref{sec:sum_rate_max}. Then, in the inner loop, we alternately update $\boldsymbol{\theta}$, $\boldsymbol{\vartheta}$ and $\boldsymbol{\omega}$ to solve for RIS reflection phase-shift coefficients $\boldsymbol{\theta}$. In the outer loop, the transmit precoding $\mathbf{w}_k$, the RIS reflection phase-shift coefficients $\boldsymbol{\theta}$, and the RIS reflection amplification factors $\mathbf{a}$ are iteratively optimized.
Similar to the sum-rate maximization problem, the ADMM algorithm is also utilized to solve for the non-convex problem (\ref{pr:pm_theta}) for updating $\boldsymbol{\theta}$, and thus the convergence cannot be mathematically proved. Simulation results in Sec. \ref{simulation} present that the proposed algorithm converges within a few iterations.
Furthermore, we give a brief computational complexity analysis of Algorithm 2. The optimizations of $\mathbf{w}_k$ and $\mathbf{a}$ are both SOCP problems and have the complexity of approximately $\mathcal{O}((KN)^3K^{1.5})$ and $\mathcal{O}(\sqrt{2K+L}KL^3)$, respectively. Similar to Algorithm 1, the computational complexity of designing $\boldsymbol{\theta}$ also includes three parts: The updates of $\boldsymbol{\theta}$, $\boldsymbol{\vartheta}$ and $\boldsymbol{\omega}$, whose complexities are approximately of order $\mathcal{O}(\sqrt{M(K+1)}MK(1+M)^3)$, $\mathcal{O}(M)$ and $\mathcal{O}(M)$, respectively. Therefore, the overall complexity of Algorithm \ref{Algorithm for power} is of order $\mathcal{O}(I_\mathrm{o}[(KN)^3K^{1.5} + I_{\mathrm{i},s}(\sqrt{M(K+1)}MK(1+M)^3) + \sqrt{2K+L}KL^3])$.

\section{Simulation Results}
\label{simulation}

\begin{figure}[t]
\centering
  \includegraphics[width = 3.0 in]{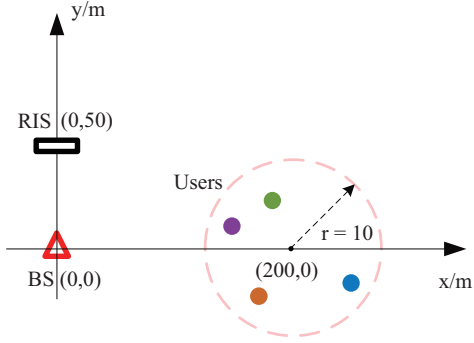}
  \caption{An illustration of the position of BS, RIS, and users.}
  \vspace{-0.4cm}
  \label{fig:position}
\end{figure}

In this section, we demonstrate extensive simulation results to verify the advantage of the sub-connected active RIS architecture and the effectiveness of the proposed joint beamforming design algorithms. The following settings are assumed throughout our simulations if not otherwise specified, which are very typical in the field of RIS \cite{Zhou}-\cite{Wu5}, \cite{Long}, \cite{Zhang}, \cite{You}-\cite{KLiu}. We assume that the BS equipped with $N = 16$ antennas is located at the origin to serve $K = 4$ downlink users which are randomly distributed within a circle with its center at ($x = 200$m, 0) and radius of 10m, as shown in Fig. \ref{fig:position}. Besides, a sub-connected active RIS consists of $M = 256$ elements is positioned at (0, $50$m) to assist the MU-MISO communication. We adopt the popular large-scale fading path-loss (PL) model: $PL(d) = C_0(d_0/d)^{\iota}$, where $C_0 = -$30dB, $d_0$ = 1m, $d$ is the distance of link and $\iota$ represents the path-loss exponent. More specifically, we set the path-loss exponents for the BS-user channels $\mathbf{h}_{\mathrm{d},k}$, BS-RIS channel $\mathbf{G}$, and RIS-user channels $\mathbf{h}_{\mathrm{r},k}$ as $\iota_{\mathrm{d},k} = 3.8$, $\iota_{\mathrm{G}} = 2.5$, and $\iota_{\mathrm{r},k} = 2.8$, $\forall k$, respectively.
In addition, the channels of BS-RIS link, BS-user links and RIS-user links are assumed to follow the Rayleigh fading model.

Furthermore, we set the dissipated power of the BS as $W_\mathrm{BS} = 6$dB \cite{KLiu}. The hardware static powers of each phase-shifter circuit and reflection amplifier are set as $W_\mathrm{PS} = 7$dBm \cite{Bjornson} and $W_\mathrm{PA} = 7$dBm \cite{Landsberg}, respectively. The total power budget for the active RIS is $P_\mathrm{RIS}^{\mathrm{tot}} = 4.15$dB.
The energy conversion efficiency is the same for the BS and the active RIS, i.e., $\nu_{1} = \nu_{2} = 1/1.1$. For simplicity, we assume the QoS requirements and the noise powers are the same for all users, i.e., $\Gamma_k = \Gamma, ~\forall k$ and $\sigma_k^2 = \sigma^2 = -80$dBm,$~\forall k$. The dynamic noise power introduced by the active RIS is set to $\sigma_{z}^2 = -80$dBm. Moreover. the penalty parameter is set as $\rho = 1$ \cite{Boyd}, \cite{HuangK}. To better verify the superiority of sub-connected active RIS structure and the proposed algorithms, we add the performance of existing fully-connected active RIS scheme \cite{Zhang} in the simulation results for comparison.

\subsection{Sum-rate Maximization Problem}

\begin{figure}
  \begin{minipage}[t]{0.48\textwidth}
     \begin{minipage}[t]{0.5\linewidth}
        \centering
        \includegraphics[width=\textwidth]{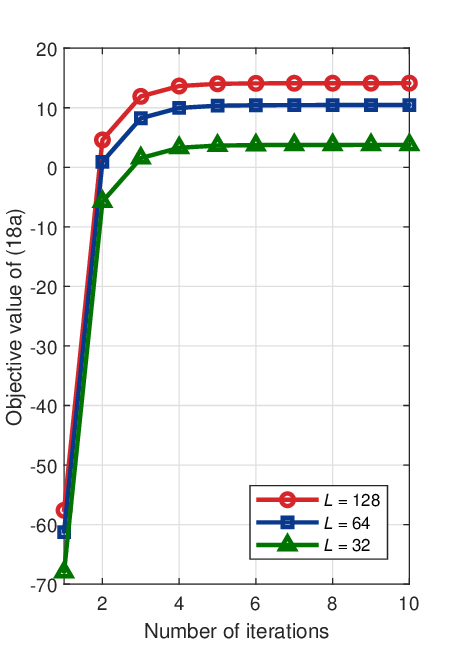}
        \centerline{(a) First inner loop.}
    \end{minipage}%
    \begin{minipage}[t]{0.5\linewidth}
        \centering
        \includegraphics[width=\textwidth]{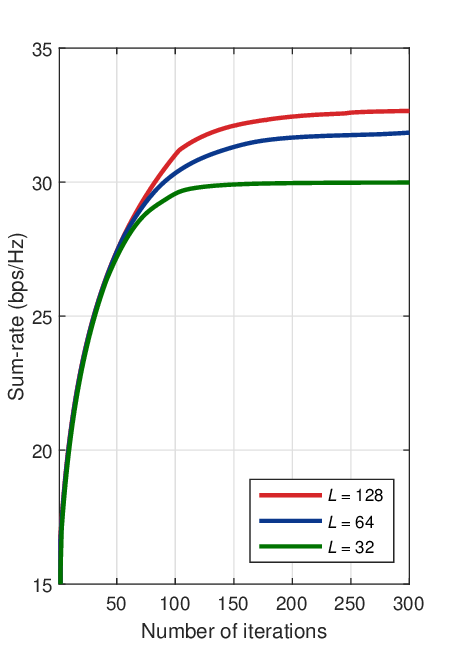}
        \centerline{(b) Outer loop.}
    \end{minipage}
    \caption{Convergence of Algorithm \ref{Algorithm for sum-rate} ($P_\mathrm{BS} = 40$dBm, $P_\mathrm{RIS}^{\mathrm{tot}} = 4.15$dB, $M = 256$, $N_\mathrm{t} = 16$, and $K = 4$).}
  \label{fig:sum-rate_max_con}
\end{minipage}
\vspace{-0.4cm}
\end{figure}

\begin{figure}[t]
    \centering
    \includegraphics[width=3.4 in]{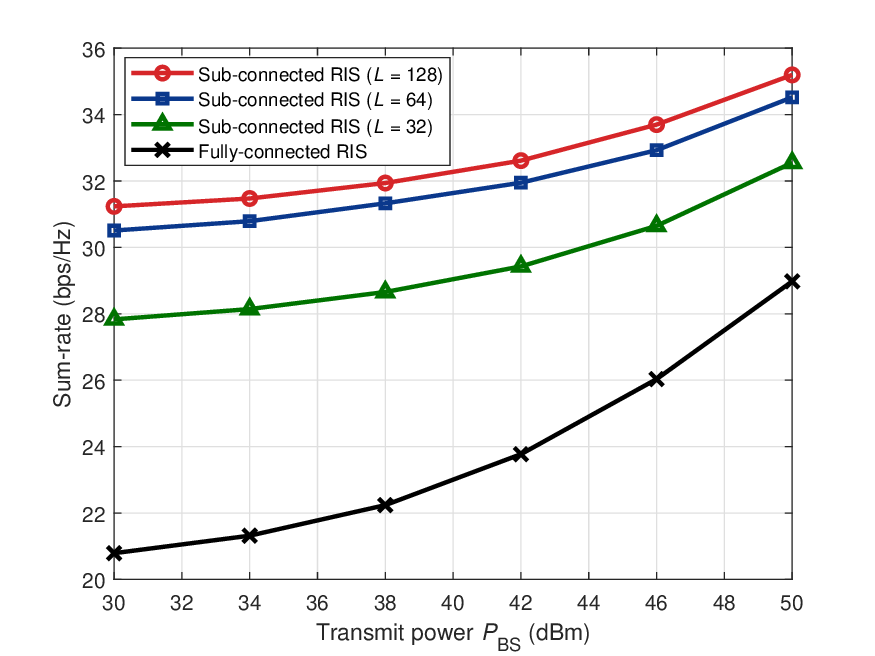}
    \caption{Sum-rate versus the transmit power $P_\mathrm{BS}$ ($P_\mathrm{RIS}^{\mathrm{tot}} = 4.15$dB, $M = 256$, $N_\mathrm{t} = 16$, and $K = 4$).}
    \label{fig:sum-rate_max_Pbs}
    \vspace{-0.4cm}
\end{figure}

In this subsection, the simulation results for the sum-rate maximization problem are demonstrated. The convergence performance of the proposed Algorithm \ref{Algorithm for sum-rate} is illustrated in Fig. \ref{fig:sum-rate_max_con}. Particularly, the convergence of the first inner loop is illustrated in Fig. \ref{fig:sum-rate_max_con}(a) and the convergence of the outer loop is presented in Fig. \ref{fig:sum-rate_max_con}(b). We can observe that the proposed algorithm exhibits satisfactory convergence performance under different sub-connected structures.

\begin{figure}[t]
    \centering
    \includegraphics[width =3.4 in]{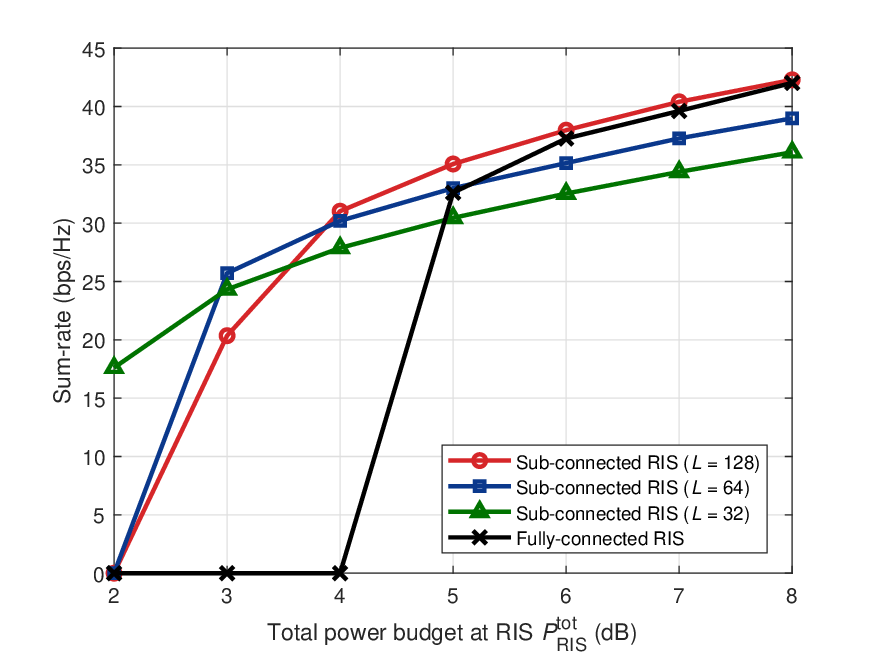}
    \caption{Sum-rate versus the power budget at RIS $P_{\mathrm{RIS}}^{\mathrm{tot}}$ ($P_\mathrm{BS} = 30$dBm, $M = 256$, $N_\mathrm{t} = 16$, and $K = 4$).}
    \label{fig:sum-rate_max_Pris}
\vspace{-0.4cm}
\end{figure}

\begin{figure}[t]
    \centering
    \includegraphics[width=3.4 in]{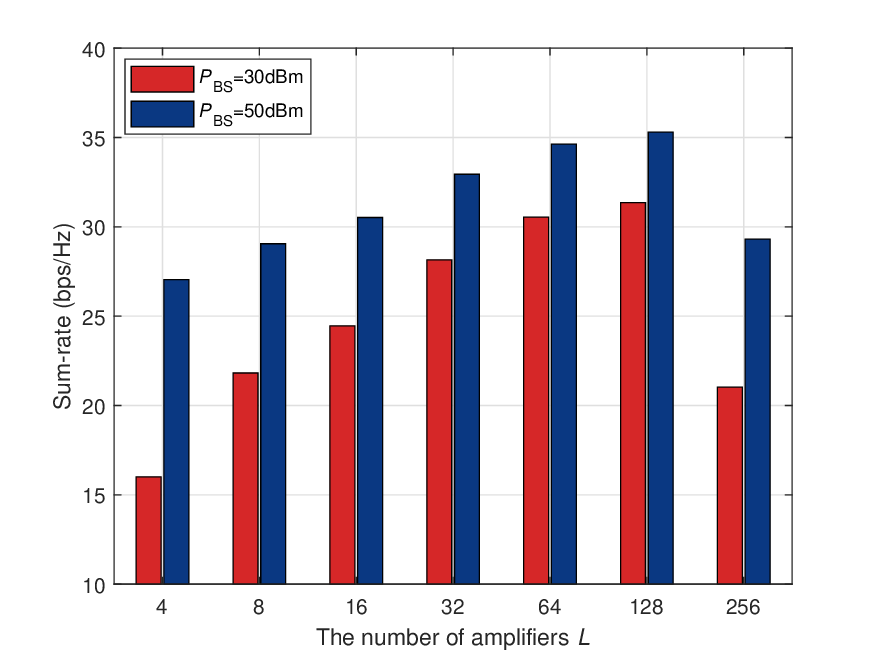}
    \caption{Sum-rate versus the number of amplifiers $L$ ($P_\mathrm{RIS}^{\mathrm{tot}} = 4.15$dB, $M = 256$, $N_\mathrm{t} = 16$, and $K = 4$).}
    \label{fig:sum-rate_max_L}
\vspace{-0.4cm}
\end{figure}

Fig. \ref{fig:sum-rate_max_Pbs} demonstrates the sum-rate performance versus the transmit power $P_\mathrm{BS}$. We can easily observe that more transmit power provides a higher sum-rate for all scenarios and the sub-connected active RIS consistently outperforms the fully-connected active RIS. Specifically, when $P_\mathrm{BS} = 30$dBm, the sub-connected scheme with $L = 64$ amplifiers can achieve up to 47\% sum-rate performance improvement compared to the fully-connected one, but employs only a quarter of the number of amplifiers. This result verifies the advancement in the hardware efficiency and performance improvement of the sub-connected structure.

The sum-rate performance versus the total power budget at the RIS $P_\mathrm{RIS}^\mathrm{tot}$ is illustrated in Fig. \ref{fig:sum-rate_max_Pris}. We can observe that when $P_\mathrm{RIS} ^\mathrm{tot}$ is relatively small, the sub-connected scheme is significantly superior to the fully-connected one, which has zero sum-rate when $P_\mathrm{RIS}^\mathrm{tot}$ is less than 4dB. This phenomenon is because the fully-connected structure consumes all the power for maintaining the operation of massive active components when $P_\mathrm{RIS}^\mathrm{tot}$ is small and no power is available for signal amplification. This result verifies the superiority of the proposed sub-connected active RIS when the power budget of RIS is limited. Even when $P_\mathrm{RIS}^{\mathrm{tot}}$ is up to 8dB, the proposed sub-connected scheme with $L = 128$ still achieves the same level of performance as the fully-connected one. In addition, Fig. \ref{fig:sum-rate_max_Pris} also shows that the optimal number of amplifiers is varying for different $P_\mathrm{RIS}^\mathrm{tot}$ cases, which is a trade-off between reflection power and design DoFs.

\begin{figure}[t]
    \centering
    \includegraphics[width =3.4 in]{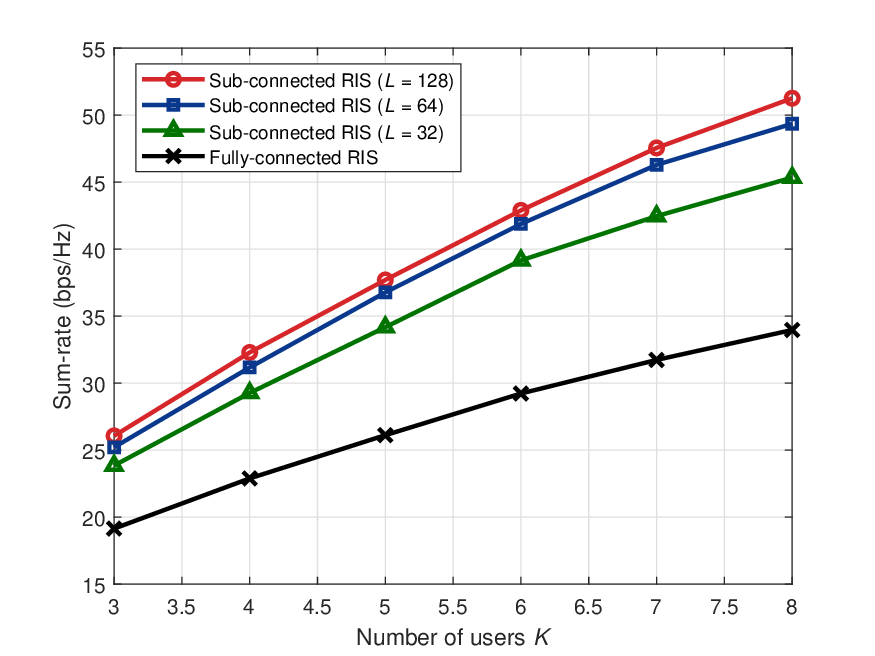}
    \caption{Sum-rate versus the number of users $K$ ($P_\mathrm{BS} = 40$dBm, $P_\mathrm{RIS}^{\mathrm{tot}} = 4.15$dB, $M = 256$, and $N_\mathrm{t} = 16$).}
    \label{fig:sum-rate_max_K}
\vspace{-0.4cm}
\end{figure}

\begin{figure}[t]
    \centering
    \includegraphics[width=3.4 in]{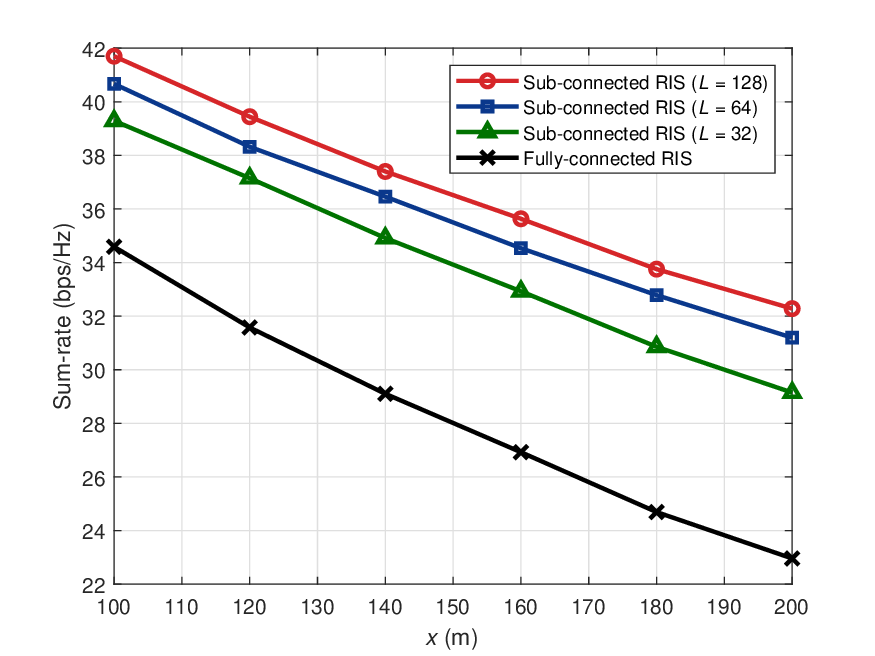}
    \caption{Sum-rate versus the location of users ($P_\mathrm{BS} = 40$dBm, $P_\mathrm{RIS}^{\mathrm{tot}} = 4.15$dB, $M = 256$, $N_\mathrm{t} = 16$, and $K = 4$).}
    \label{fig:sum-rate_max_dis}
    \vspace{-0.4cm}
\end{figure}

Fig. \ref{fig:sum-rate_max_L} depicts the sum-rate performance versus the number of amplifiers $L$, where $L = 256$ actually represents the traditional fully-connected structure. It can be seen that the sum-rate of proposed sub-connected active RIS-assisted system first increases with the growth of $L$ and then decreases, which achieves a maximum when $L = 128$ for both $P_\mathrm{BS} = 30$dBm and $P_\mathrm{BS} = 50$dBm cases. This interesting phenomenon is because that when the total power budget at the RIS is limited, the reduction in the number of amplifiers will allocate more power for signal amplification. Nevertheless, it also results in the decrease of DoFs for the beamforming design and leads to performance degradation with too few amplifiers. Therefore, it is very crucial to select an appropriate grouping strategy (i.e., the number of amplifiers) to balance the hardware cost, power consumption, and sum-rate performance. As illustrated in Fig. \ref{fig:sum-rate_max_L}, when $L = 16/32/64/128$, the proposed sub-connected structure achieves better performance than the compared fully-connected scheme, which verifies the advancement of sub-connected active RIS. Another important conclusion we can draw from Fig. \ref{fig:sum-rate_max_L} is that the sub-connected architecture can achieve similar sum-rate performance as the fully-connected scheme with only about 3\% of the number of amplifiers (i.e., $L = 8$), which is a dramatic hardware cost reduction.

\begin{figure}
\vspace{-0.1cm}
  \begin{minipage}[t]{0.48\textwidth}
     \begin{minipage}[t]{0.50\linewidth}
        \centering
        \includegraphics[width=\textwidth]{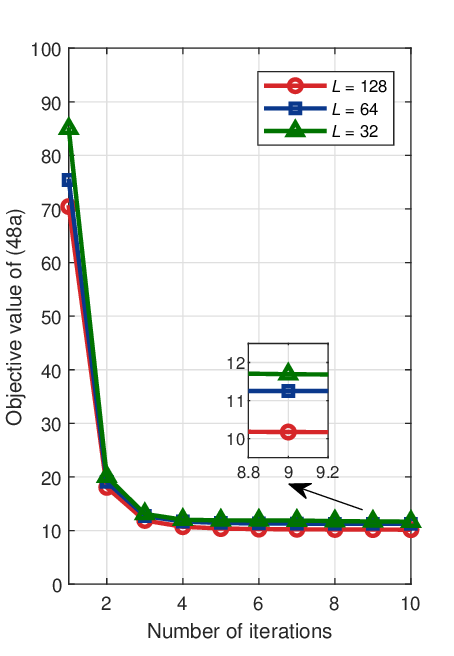}
        \centerline{(a) First inner loop.}
    \end{minipage}%
    \begin{minipage}[t]{0.50\linewidth}
        \centering
        \includegraphics[width=\textwidth]{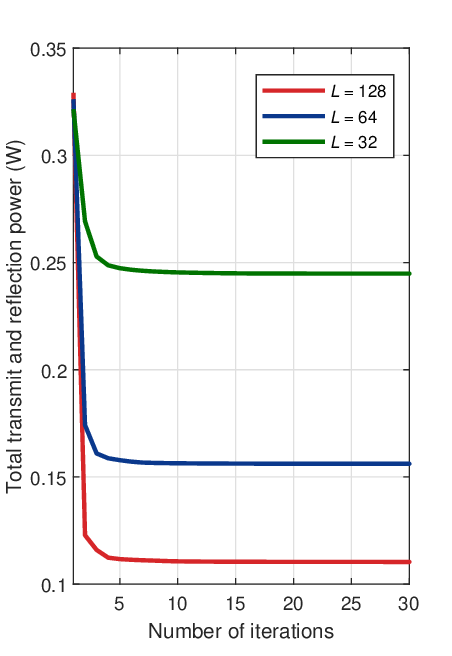}
        \centerline{(b) Outer loop.}
    \end{minipage}
    \caption{Convergence of Algorithm \ref{Algorithm for power} ($\Gamma = 8$dB, $M = 256$, $N_\mathrm{t} = 16$, and $K = 4$).}
  \label{fig:power_min_con}
\end{minipage}
\vspace{-0.4cm}
\end{figure}

In addition, we present the sum-rate versus the number of users in Fig. \ref{fig:sum-rate_max_K}. It can be observed that with the growth of the number of users, the sum-rate of both sub-connected and fully-connected active RIS-aided systems increases owing to more multi-user diversity. Moreover, it further verifies that the proposed sub-connected scheme always outperforms the fully-connected one and the performance gap becomes more significant with larger $K$. In Fig. \ref{fig:sum-rate_max_dis}, we shows the sum-rate performance versus the location of users. Not surprisingly, when the users are away from the BS and RIS (i.e., $x$ becomes larger), the path-loss increases and thereby the sum-rate of both two schemes decreases. The sub-connected active RIS can always outperform the fully-connected one under different user location cases.

\subsection{Power Minimization problem}

In this subsection, we present the simulation results for the power minimization problem. Firstly, the convergence of Algorithm \ref{Algorithm for power} is shown in Fig. \ref{fig:power_min_con}. Obviously, the convergences of both the inner and outer loops are very rapid.

\begin{figure}[t]
\centering
\includegraphics[width=3.4 in]{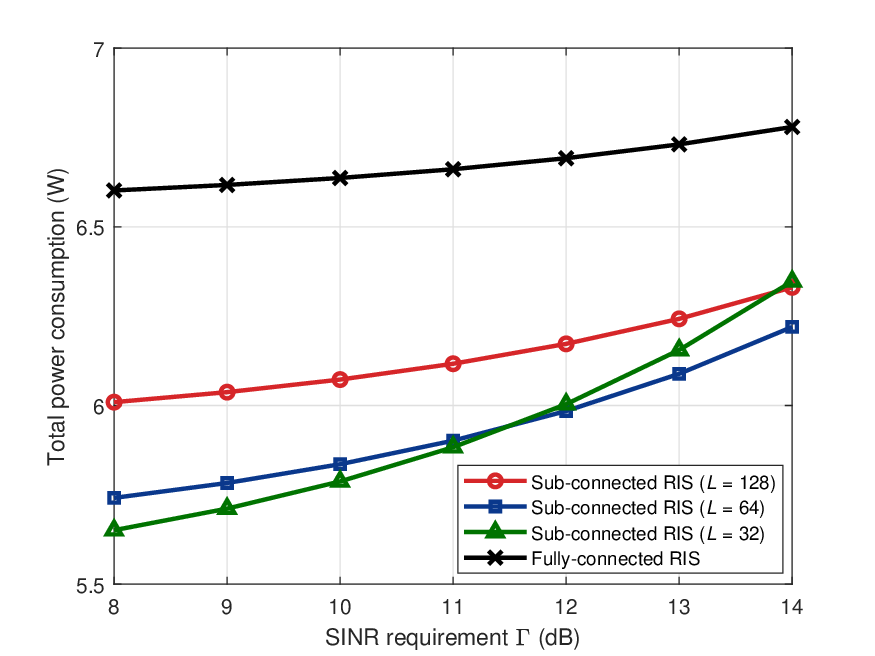}
\vspace{0.25cm}
\caption{Total power consumption versus SINR requirement $\Gamma$ ($M = 256$, $N_\mathrm{t} = 16$, and $K = 4$).}
\label{fig:power_min_Gamma}
\vspace{-0.4cm}
\end{figure}%

\begin{figure}
    \centering
    \includegraphics[width=3.4 in]{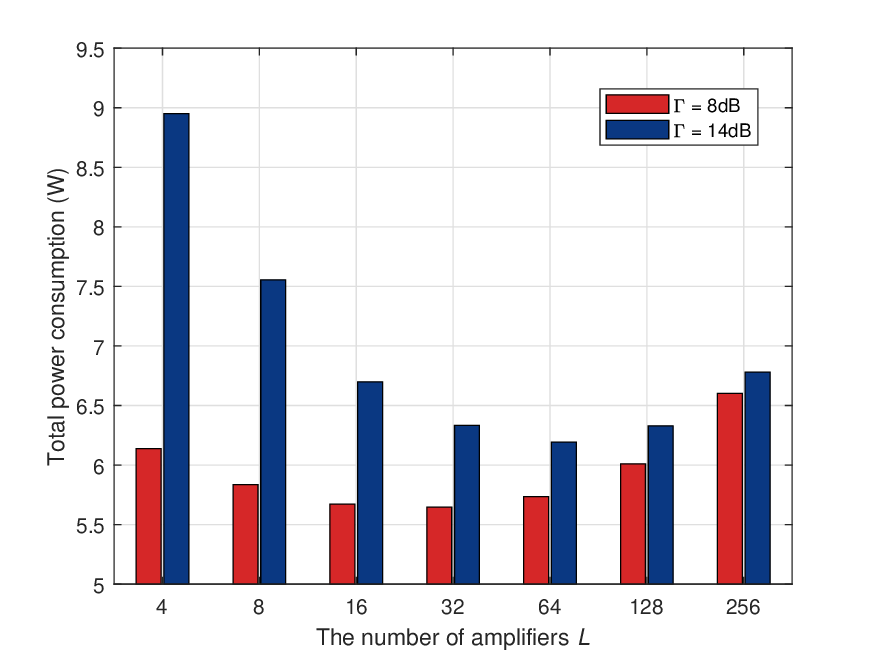}
    \caption{Total power consumption versus the number of amplifiers $L$ ($M = 256$, $N_\mathrm{t} = 16$, and $K = 4$).}
    \label{fig:power_min_L}
\vspace{-0.4cm}
\end{figure}

The total power consumption versus the users' SINR requirement $\Gamma$ is plotted in Fig. \ref{fig:power_min_Gamma}. Compared to the conventional fully-connected scheme, our proposed sub-connected architecture with $L = 32$ needs only about 85\% power and 1/8 amplifiers to meet the same $\Gamma = 8$dB requirement, which validates the advantage in energy efficiency and hardware efficiency of the sub-connected active RIS structure. More interestingly, the total power consumption of a sub-connected scheme with fewer amplifiers grows at a faster rate as the SINR requirement grows. This suggests that the increase in transmit/reflection power to satisfy QoS constraints is dominant compared to the decrease in hardware power consumption at a higher SINR requirement. Therefore, based on the above analysis, $L=64$ is an appropriate grouping strategy for various QoS requirements.

In Fig. \ref{fig:power_min_L}, we display the total power consumption versus the number of amplifiers $L$. Again, the case with $L = 256$ represents the fully-connected structure. We can notice that with a relatively small users' SINR requirement (e.g., $\Gamma = 8$dB), the sub-connected active RIS-aided system consumes less power in comparison to the fully-connected active RIS-aided system for all cases of $L$. Moreover, when the SINR requirement is large (e.g., $\Gamma = 14$dB), the performance degradation caused by fewer design DoFs becomes more significant. For different SINR requirements, the sub-connected structure with $L=16/32/64/128$ is always more hardware and power efficient compared to the fully-connected structure. Unlike the sum-rate maximization problem where $L=128$ is the optimal grouping strategy as illustrated in Fig. \ref{fig:sum-rate_max_L}, there are different solutions under different SINR constraints, e.g., $L = 32$ and $L = 64$ are the best choices for $\Gamma = 8$dB and $\Gamma = 14$dB cases, respectively, which is a balance between transmit/reflection power consumption and hardware power consumption.

\begin{figure}[t]
\centering
\includegraphics[width=3.4 in]{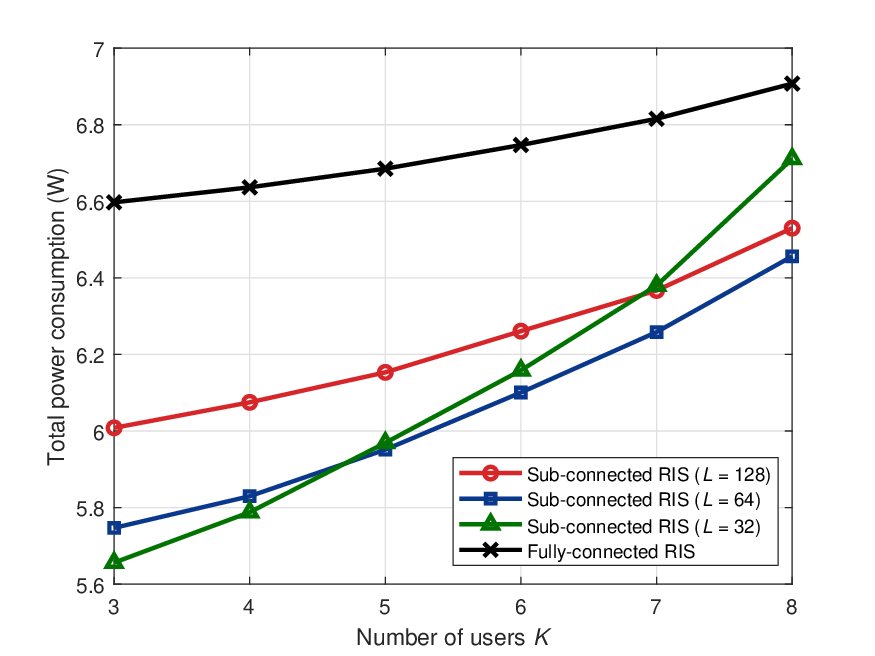}
\caption{Total power consumption versus the number of users $K$ ($\Gamma = 10$dB, $M = 256$, and $N_\mathrm{t} = 16$).}
\label{fig:power_min_K}
\vspace{-0.4cm}
\end{figure}

\begin{figure}
    \centering
    \includegraphics[width=3.4 in]{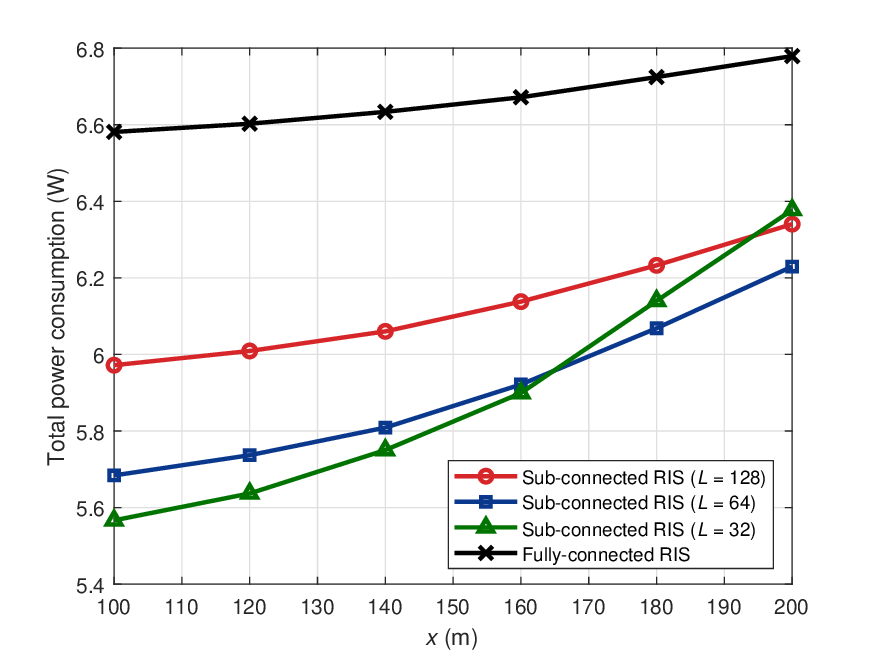}
    \caption{Total power consumption versus the location of users $x$ ($\Gamma = 14$dB, $M = 256$, $N_\mathrm{t} = 16$, and $K = 4$).}
    \label{fig:power_min_dis}
\vspace{-0.4cm}
\end{figure}

Finally, Figs. \ref{fig:power_min_K} and \ref{fig:power_min_dis} illustrate the total power consumption versus the number and location of users, respectively. Obviously, the proposed sub-connected scheme can always achieve satisfactory performance than its competitor, i.e., the fully-connected scheme. As expected, the increase in the number of serving users and the horizontal distance $x$ between the BS and users' center will lead to the growth of the total power consumption of the system.

\section{Conclusions}
In this paper, we considered the recently emerging sub-connected array architecture of active RIS to reduce the power consumption and cost of active components. After providing a reformative and accurate signal model of the sub-connected active RIS, we investigated the joint designs of transmit precoding and reflection beamforming for both sum-rate maximization problem and power minimization problem in MU-MISO systems. Based on the FP, BCD, ADMM, MM, and SOCP methods, we successfully developed efficient solutions to optimize the challenging non-convex problems. Simulation results demonstrated the effectiveness of the proposed algorithms and also confirmed that the sub-connected active RIS is superior to the traditional fully-connected structure in both hardware cost and energy-saving perspectives.

Although we have confirmed the superiority of the novel sub-connected active RIS, it also introduces new challenges and difficulties, e.g., dynamic power allocation, dependency between incident signal power and amplification factor, optimal grouping strategy, energy efficiency analysis, channel estimation, hardware implementation, etc. Based on this initial work, we will further investigate these issues in our future studies.


\begin{thebibliography}{99}
\bibitem{SAM} Q. Zhu, M. Li, Y. Liu, and Q. Liu, ``Joint beamforming design for sub-connected active reconfigurable intelligent surface," in \textit{Proc. IEEE Sensor Array Multichannel Signal Process. Workshop (SAM)}, Trondheim, Norway, Jun. 2022.
\bibitem{Pan} C. Pan \textit{et al.}, ``Reconfigurable intelligent surfaces for 6G systems: Principles, applications, and research directions," \textit{IEEE Commun. Mag.}, vol. 59, no. 6, pp. 14-20, Jun. 2021.
\bibitem{GC}G. C. Alexandropoulos, G. Lerosey, M. Debbah, and M. Fink, ``Reconfigurable intelligent surfaces and metamaterials: The potential of wave propagation control for 6G wireless communications," Jun. 2020. [Online]. Available: http://arxiv.org/abs/2006.11136
\bibitem{Basar} E. Basar, M. Di Renzo, J. De Rosny, M. Debbah, M.-S. Alouini, and R. Zhang, ``Wireless communications through reconfigurable intelligent surfaces," \textit{IEEE Access}, vol. 7, pp. 116753-116773, Aug. 2019.
\bibitem{Renzo}M. Di Renzo \textit{et al.}, ``Smart radio environments empowered by reconfigurable intelligent surfaces: How it works, state of research, and the road ahead," \textit{IEEE J. Sel. Areas Commun.}, vol. 38, no. 11, pp. 2450-2525, Nov. 2020.
\bibitem{Wu2} Q. Wu and R. Zhang, ``Towards smart and reconfigurable environment: Intelligent reflecting surface aided wireless network," \textit{IEEE Commun. Mag.}, vol. 58, no. 1, pp. 106-112, Jan. 2020.
\bibitem{Wu} Q. Wu, S. Zhang, B. Xiong, C. You, and R. Zhang, ``Intelligent reflecting surface aided wireless communications: A tutorial," \textit{IEEE Trans. Commun.}, vol. 69, no. 5, pp. 3313-3351, May 2021.
\bibitem{Zhou}S. Zhou, W. Xu, K. Wang, M. D. Renzo, and M. Alouini, ``Spectral and energy efficiency of IRS-assisted MISO communication with hardware impairments," \textit{IEEE Wireless Commun. Lett.}, vol. 9, no. 9, pp. 1366-1369, Sep. 2020.
\bibitem{Huang}C. Huang, A. Zappone, G. C. Alexandropoulos, M. Debbah, and C. Yuen, ``Reconfigurable intelligent surfaces for energy efficiency in wireless communication," \textit{IEEE Trans. Wireless Commun.}, vol. 18, no. 8, pp. 4157-4170, Aug. 2019.
\bibitem{Zhou2}G. Zhou, C. Pan, H. Ren, K. Wang, and A. Nallanathan, ``Intelligent reflecting surface aided multigroup multicast MISO communication systems," \textit{IEEE Trans. Signal Process.}, vol. 68, pp. 3236-3251, Apr. 2020.
\bibitem{WangP} H. Guo, Y. Liang, J. Chen, and E. G. Larsson, ``Weighted sum-rate maximization for reconfigurable intelligent surface aided wireless networks," \textit{IEEE Trans. Wireless Commun.}, vol. 19, no. 5, pp. 3064-3076, May 2020.
\bibitem{Wu3} Q. Wu and R. Zhang, ``Intelligent reflecting surface enhanced wireless network via joint active and passive beamforming design," \textit{IEEE Trans. Wireless Commun.}, vol. 18, no. 11, pp. 5394-5409, Nov. 2019.
\bibitem{Wu4} Q. Wu and R. Zhang, ``Beamforming optimization for wireless network aided by intelligent reflecting surface with discrete phase shifts," \textit{IEEE Trans. Commun.}, vol. 68, no. 3, pp. 1838-1851, Mar. 2020.
\bibitem{CuiM}M. Cui, G. Zhang, and R. Zhang, ``Secure wireless communication via intelligent reflecting surface," \textit{IEEE Wireless Commun. Lett.}, vol. 8, no. 5, pp. 1410-1414, Oct. 2019.
\bibitem{Wu5} Q. Wu and R. Zhang, ``Joint active and passive beamforming optimization for intelligent reflecting surface assisted SWIPT under QoS constraints," \textit{IEEE J. Sel. Areas Commun.}, vol. 38, no. 8, pp. 1735-1748, Aug. 2020.
\bibitem{Li2} H. Li, W. Cai, Y. Liu, M. Li, Q. Liu, and Q. Wu, ``Intelligent reflecting surface enhanced wideband MIMO-OFDM communications: From practical model to reflection optimization," \textit{IEEE Trans. Commun.}, vol. 69, no. 7, pp. 4807-4820, Jul. 2021.
\bibitem{LiuR} R. Liu, M. Li, Q. Liu, and A. Lee Swindlehurst, ``Joint symbol-level precoding and reflecting designs for IRS-enhanced MU-MISO systems," \textit{IEEE Trans. Wireless Commun.}, vol. 20, no. 2, pp. 798-811, Feb. 2021.
\bibitem{LiuR2} R. Liu, M. Li, Qian Liu, and A. Lee Swindlehurst, ``Intelligent reflecting surface based passive information transmission: A symbol-level precoding approach," \textit{IEEE Trans. Veh. Technol.}, vol. 70, no. 7, pp. 6735-6749, Jul. 2021.
\bibitem{WangX} X. Wang, Z. Fei, J. Huang, and H. Yu, ``Joint waveform and discrete phase shift design for RIS-assisted integrated sensing and communication system under Cram\'{e}r-Rao bound constraint," \textit{IEEE Trans. Veh. Technol.}, vol. 71, no. 1, pp. 1004-1009, Jan. 2022.
\bibitem{LiuR3} R. Liu, M. Li, Y. Liu, Q. Wu, and Q. Liu, ``Joint transmit waveform and passive beamforming design for RIS-aided DFRC systems," \textit{IEEE J. Sel. Topics Signal Process.}, to appear, DOI: 10.1109/JSTSP.2022.3172788.

\bibitem{Long}R. Long, Y.-C. Liang, Y. Pei, and E. G. Larsson, ``Active reconfigurable intelligent surface-aided wireless communications," \textit{IEEE Trans. Wireless Commun.}, vol. 20, no. 8, pp. 4962-4975, Aug. 2021.
\bibitem{Zhang}Z. Zhang, L. Dai, X. Chen, C. Liu, F. Yang, R. Schober, and H. V. Poor, ``Active RIS vs. passive RIS: Which will prevail in 6G?" Mar. 2022. [Online]. Available: https://arxiv.org/abs/2103.15154

\bibitem{HuC} C. Hu, L. Dai, S. Han, and X. Wang, ``Two-timescale channel estimation for reconfigurable intelligent surface aided wireless communications," \textit{IEEE Trans. Commun.}, vol. 69, no. 11, pp. 7736-7747, Nov. 2021.
\bibitem{PanC} C. Pan, H. Ren, K. Wang, W. Xu, M. Elkashlan, A. Nallanathan, and L. Hanzo, ``Multicell MIMO communications relying on intelligent reflecting surfaces," \textit{IEEE Trans. Wireless Commun.}, vol. 19, no. 8, pp. 5218-5233, Aug. 2020.
\bibitem{Najafi} M. Najafi, V. Jamali, R. Schober, and H. V. Poor, ``Physics-based modeling and scalable optimization of large intelligent reflecting surfaces," \textit{IEEE Trans. Commun.}, vol. 69, no. 4, pp. 2673-2691, Apr. 2021.



\bibitem{Bousquet}J. Bousquet, S. Magierowski, and G. G. Messier, ``A 4-GHz active scatterer in 130-nm CMOS for phase sweep amplify-and-forward," \textit{IEEE Trans. Circuits Syst. I}, vol. 59, no. 3, pp. 529-540, Mar. 2012.
\bibitem{Landsberg} N. Landsberg and E. Socher, ``A low-power 28-nm CMOS FD-SOI reflection amplifier for an active F-band reflectarray," \textit{IEEE Trans. Microw. Theory Techn.}, vol. 65, no. 10, pp. 3910-3921, May 2017.
\bibitem{KuoC} C.-N. Kuo, Y.-H. Liu, and R.-H. Gao, ``A 57-GHz CMOS reflection amplifier in 90-nm CMOS," \textit{IEEE Microw. Wireless Compon. Lett.}, vol. 32, no. 4, pp. 335-338, Apr. 2022.

\bibitem{You}C. You and R. Zhang, ``Wireless communication aided by intelligent reflecting surface: Active or passive?" \textit{IEEE Wireless Commun. Lett.}, vol. 10, no. 12, pp. 2659-2663, Dec. 2021.
\bibitem{Khoshafa}M. H. Khoshafa, T. M. N. Ngatched, M. H. Ahmed, and A. R. Ndjiongue, ``Active reconfigurable intelligent surfaces-aided wireless communication system," \textit{IEEE Commun. Lett.}, vol. 25, no. 11, pp. 3699-3703, Nov. 2021.
\bibitem{Dong}L. Dong, H.-M. Wang and J. Bai, ``Active reconfigurable intelligent surface aided secure transmission," \textit{IEEE Trans. Veh. Technol.}, vol. 71, no. 2, pp. 2181-2186, Feb. 2022.
\bibitem{GaoY} Y. Gao, Q. Wu, G. Zhang, W. Chen, D. W. K. Ng, and M. D. Renzo, ``Beamforming optimization for active intelligent reflecting surface-aided SWIPT," Mar. 2022. [Online]. Available: https://arxiv.org/abs/2203.16093
\bibitem{ZengP} P. Zeng, D. Qiao, Q. Wu, and Y. Wu, ``Throughput maximization for active intelligent reflecting surface aided wireless powered communications," \textit{IEEE Wireless Commun. Lett.}, vol. 11, no. 5, pp. 992-996, May 2022.
\bibitem{GeY}Y. Ge and J. Fan, ``Active intelligent reflecting surface assisted secure air-to-ground communication with UAV jittering," Mar. 2022. [Online]. Available: https://arxiv.org/abs/2203.12296

\bibitem{Nguyen} N. T. Nguyen, V.-D. Nguyen, Q. Wu, A. Tolli, S. Chatzinotas, and M. Juntti, ``Hybrid active-passive reconfigurable intelligent surface-assisted multi-user MISO systems," Mar. 2022. [Online]. Available: https://arxiv.org/abs/2203.07042
\bibitem{Nguyen2} N. T. Nguyen, D. Vu, K. Lee and M. Juntti, ``Hybrid relay-reflecting intelligent surface-assisted wireless communications," \textit{IEEE Trans. Veh. Technol.}, to appear, DOI: 10.1109/TVT.2022.3158686.
\bibitem{KLiu} K. Liu, Z. Zhang, L. Dai, S. Xu, and F. Yang, ``Active reconfigurable intelligent surface: Fully-connected or sub-connected?" \textit{IEEE Commun. Lett.}, vol. 26, no. 1, pp. 167-171, Jan. 2022.
\bibitem{MaoZ}Z. Mao, W. Wang, Q. Xia, C. Zhong, X. Pan, and Z. Ye, ``Element-grouping intelligent reflecting surface: Electromagnetic-compliant model and geometry-based optimization," \textit{IEEE Trans. Wireless Commun.}, vol. 21, no. 7, pp. 5362-5376, Jul. 2022.
\bibitem{Kundu} N. K. Kundu, Z. Li, J. Rao, S. Shen, M. R. McKay, and R. Murch, ``Optimal grouping strategy for reconfigurable intelligent surface assisted wireless communications," \textit{IEEE Wireless Commun. Lett.}, vol. 11, no. 5, pp. 1082-1086, May 2022.
\bibitem{YanY} Y. Yang, B. Zheng, S. Zhang, and R. Zhang, ``Intelligent reflecting surface meets OFDM: Protocol design and rate maximization," \textit{IEEE Trans. Commun.}, vol. 68, no. 7, pp. 4522-4535, Jul. 2020.
\bibitem{Tasci} R. A. Tasci, F. Kilinc, E. Basar, and G. C. Alexandropoulos, ``A new RIS architecture with a single power amplifier: Energy efficiency and error performance analysis," \textit{IEEE Access}, vol. 10, pp. 44804-44815, Apr. 2022.
\bibitem{ZhengB} B. Zheng, C. You, W. Mei, and R. Zhang, ``A survey on channel estimation and practical passive beamforming design for intelligent reflecting surface aided wireless communications," \textit{IEEE Commun. Surv. Tut.}, vol. 24, no. 2, pp. 1035-1071, Feb. 2022.
\bibitem{Swindlehurst} A. L. Swindlehurst, G. Zhou, R. Liu, C. Pan, and M. Li, ``Channel estimation with reconfigurable intelligent surfaces - A general framework," \textit{in Proc. IEEE}, to appear, DOI: 10.1109/JPROC.2022.3170358.
\bibitem{WangZ} Z. Wang, L. Liu, and S. Cui, ``Channel estimation for intelligent reflecting surface assisted multiuser communications: Framework, algorithms, and analysis," \textit{IEEE Trans. Wireless Commun.}, vol. 19, no. 10, pp. 6607-6620, Oct. 2020.

\bibitem{ShenK} K. Shen and W. Yu, ``Fractional programming for communication systems-Part \uppercase\expandafter{\romannumeral2}: Uplink scheduling via matching," \textit{IEEE Trans. Signal Process.}, vol. 66, no. 10, pp. 2631-2644, May 2018.
\bibitem{Grant}M. Grant and S. Boyd, ``CVX: MATLAB software for disciplined convex programming," 2016. [Online]. Available: http://cvxr.com/cvx
\bibitem{Boyd}S. Boyd, N. Parikh, E. Chu, B. Peleato, and J. Eckstein, ``Distributed optimization and statistical learning via the alternating direction method of multipliers," \textit{Found. Trends Mach. Learn.}, vol. 3, no. 1, pp. 1-122, Jan. 2011.
\bibitem{SunY}Y. Sun, P. Babu, and D. P. Palomar, ``Majorization-minimization algorithms in signal processing, communications, and machine learning," \textit{IEEE Trans. Signal Process.}, vol. 65, no. 3, pp. 794-816, Feb. 2017.
\bibitem{WangKY} K.-Y. Wang, A. Man-Cho So, T.-H. Chang, W.-K. Ma, and C.-Y. Chi, ``Outage constrained robust transmit optimization for multiuser MISO downlinks: Tractable approximations by conic optimization," \textit{IEEE Trans. Signal Process.}, vol. 62, no. 21, pp. 5690-5705, Nov. 2014.
\bibitem{Absil} P.-A. Absil, R. Mahony, and R. Sepulchre, \textit{Optimization Algorithms on Matrix Manifolds}. Princeton, NJ, USA: Princeton Univ. Press, 2009.
\bibitem{Bjornson}E. Bj\"{o}rnson, \"{O}. \"{O}zdogan, and E. G. Larsson, ``Intelligent reflecting surface versus decode-and-forward: How large surfaces are needed to beat relaying?" \textit{IEEE Wireless Commun. Lett.}, vol. 9, no. 2, pp. 244-248, Oct. 2020.
\bibitem{HuangK} K. Huang and N. D. Sidiropoulos, ``Consensus-ADMM for general quadratically constrained quadratic programming," \textit{IEEE Trans. Signal Process.}, vol. 64, no. 20, pp. 5297-5310, Oct. 2016.
\end{thebibliography}
\end{document}